\documentclass[aps, pre,twocolumn, superscriptaddress,showpacs,longbibliography]{revtex4-2}
\usepackage{dcolumn}% Align table columns on decimal point
\usepackage{bm}% bold math
\usepackage{graphics}
\usepackage{epsfig}
\usepackage{amsmath,amssymb,bm}
\usepackage{graphicx}% Include nodes files
\usepackage{color}
\usepackage{amsthm}
\usepackage{amssymb}
\usepackage{verbatim}
\begin{document}

\newcommand{\gin}[1]{{\bf\color{blue}#1}}
\newcommand{\be}{\begin{equation}}
\newcommand{\ee}{\end{equation}}
\newcommand{\bea}{\begin{eqnarray}}
\newcommand{\eea}{\end{eqnarray}}

\def\bc{\begin{center}}
\def\ec{\end{center}}
\newcommand{\avg}[1]{\langle{#1}\rangle}
\newcommand{\Avg}[1]{\left\langle{#1}\right\rangle}
\newcommand{\Bavg}[1]{\Bigl\langle{#1}\Bigr\rangle}
\newcommand{\eq}[1]{(\ref{#1})}

\def\ie{\textit{i.e.}}
\def\etal{\textit{et al.}}
\def\m{\vec{m}}
\def\G{\mathcal{G}}

\title{The theory of percolation on hypergraphs}

\author{Ginestra Bianconi}

\affiliation{
School of Mathematical Sciences, Queen Mary University of London, London, E1 4NS, United Kingdom}
\affiliation{The Alan Turing Institute,  96  Euston  Road,  London,  NW1  2DB,  United  Kingdom}

\author{Sergey N. Dorogovtsev}
\affiliation{Departamento de F\'{\i}sica da Universidade de Aveiro $\&$ I3N, 3810-193, Aveiro, Portugal} 
\affiliation{Ioffe Physico-Technical Institute, 194021 St. Petersburg, Russia}

\begin{abstract}
 {Hypergraphs capture the higher-order interactions in complex systems and always admit a factor graph representation, consisting of a bipartite network of nodes and hyperedges. As hypegraphs are ubiquitous, investigating  hypergraph robustness is a problem of major research interest. In the literature the robustness of hypergraphs as been so far only treated adopting factor-graph percolation which describe well higher-order interactions which remain  functional even after  the  removal of one of more of their nodes. This approach, however, fall short to describe situations in which higher-order interactions fail when anyone of their nodes is removed, this latter scenario applying for instance to supply chains, catalytic networks, protein-interaction networks, networks of chemical reactions, etc. Here we show that in these cases the correct process to investigate is hypergraph percolation with is distinct from factor graph percolation.  We build a message-passing theory of hypergraph percolation and we investigate its critical behavior using generating function formalism supported by Monte Carlo simulations on random graph and real data. Notably, we show that the node percolation threshold on hypergraphs exceeds node percolation threshold on factor graphs. Furthermore we show that differently from what happens in ordinary graphs, on hypergraphs the node percolation threshold and hyperedge percolation threshold do not coincide, with the node percolation threshold exceeding the hyperedge percolation threshold. These results demonstrate that any fat-tailed cardinality distribution of hyperedges cannot lead to the hyper-resilience phenomenon in hypergraphs in contrast to their factor graphs, where the divergent second moment of a cardinality distribution guarantees zero percolation threshold.
}
\end{abstract}

\maketitle

%%%%%%%%%%%%%%%%%%%%%%
%%%%%%%%%%%%%%%%%%%%%%
%%%%%%%%%%%%%%%%%%%%%%
%%%%%%%%%%%%%%%%%%%%%%

%%%%%%%%%%%%%%%%%%%%%%
%%%%%%%%%%%%%%%%%%%%%%
%%%%%%%%%%%%%%%%%%%%%%
%%%%%%%%%%%%%%%%%%%%%%

\section{Introduction}
\label{s00}

Hypergraphs and simplicial complexes form the important class of networks---so-called higher-order networks \cite{benson2016higher,battiston2020networks,bianconi2021higher,boccaletti2023structure,torres2021and}---representing the systems of multi-node interactions. Growing research interest is addressed to both modelling \cite{bianconi2021higher,bianconi2022statistical,barthelemy2022class,krapivsky2023random}
 higher-order network structure and investigating dynamical processes on top of them. Many processes and cooperative models on the higher-order networks significantly differ from those on ordinary networks in which each edge interconnects a pair of nodes \cite{battiston2021physics,bianconi2021higher,majhi2022dynamics}. These include opinion dynamics, game theory, synchronization etc. Despite a few works have already addressed problems related to  the robustness of hypergraphs \cite{coutinho2020covering,sun2021higher,sun2023dynamic,lee2023k,peng2022disintegrate,peng2023message} 
we still lack a theory for hyperedge and node percolation on hypergraphs. 
In this work we focus on the hypergraphs. The hyperedges of a hypergraph can be treated as a second type of nodes (factor nodes), and hence the hypergraphs are equivalent to the factor graphs which are bipartite graphs 
%%between 
based on the original nodes and the factor nodes. 
This representation of hypergraphs provides one with a straightforward way to 
treating 
their structural properties \cite{sun2021higher,peng2022disintegrate} and models and processes on them, since bipartite networks are well studied. 
For example, the giant connected component of a hypergraph coincides with the giant connected component of the corresponding factor graph. Similarly, the hyperedge percolation problem for a hypergraph (removal of hyperedges) conforms the result of the removal of the factor nodes from the factor graph. 
However, there exist problems, nonequivalent on the hypergraphs and on the factor graphs. 
The origin of this difference is the distinct effect of the removal of a node from a hypergraph and the removal of a 
node from the factor graph. Indeed, the deletion of a node in a hypergraph also removes all the adjacent hyperedges; no hyperedge can change its cardinality (unless some additional rule for transformation of hyperedges is implemented). In contrast to this, the deletion of a node in a factor graph doesn't lead to the removal of factor nodes, only the connections of the neighboring factor nodes to the removed node disappear. 

 {There are two classes of higher-order interactions captured by two distinct types of hyperedges. In the first class, including networks of social interactions,  the hyperedges only break if all but one of their nodes fail. Therefore these hyperedges can sustain the failure of one and even more of their nodes.
For this class, existing theories and models treating hypergraphs as their factor graphs \cite{sun2021higher,lee2023k,mancastroppa2023hyper} work perfectly. 
However, there exists a second important class of hyperedges, which fail as soon as one of their node is damaged. In particular, this class includes supply chains and catalytic networks \cite{thurner2010schumpeterian,hanel2005phase}, protein-interaction networks \cite{klimm2021hypergraphs}, and networks of chemical reactions \cite{jost2019hypergraph}. For instance the removal of a raw material will impede the production of a product, the absence of a protein will impede the formation of a protein complex and the absence of a reactant will impede a chemical reaction to occur. For these important hypergraphs, existing theories doesn't work. 
In the present paper we develop a percolation theory for such hypergraphs and show that the difference from percolation on factor graphs can be dramatic.} 
%%, provide an example for higher-order interactions of this kind. Such systems of multi-node interactions are described by hypergraphs, in which the removal of a node results in the disappearance of all the adjacent hyperedges. For instance the removal of a raw material will impede the production of a product, the absence of a protein will impede the formation of a protein complex and the absence of a reactant will impede a chemical reaction to occur. Node percolation problems for these two classes of systems qualitatively differ from each other, compare Refs.~\cite{sun2021higher} and \cite{bianconi2023theory}, although edge percolation on hypergraphs coincides with factor node percolation on corresponding factor graphs.   

To demonstrate the difference between percolation on hypergraphs and on factor graph, we explore percolation problem for hypergraphs and compare it to percolation on factor graphs. 
Our approach builds on the message-passing theory for percolation \cite{bianconi2018multilayer,newman2023message,hartmann2006phase,karrer2010message,zhao2013antagonistic,cellai2016message,radicchi2017redundant,watanabe2014cavity,cantwell2023heterogeneous,bianconi2018rare,cantwell2019message} which is a special case to message-passing algorithms widely used also in epidemic spreading, Ising models and combinatorial optimization \cite{mezard2009information,yoon2011belief,sun2021competition,liu2011controllability,menichetti2014network,bianconi2021message,weigt2006message}
and on the fundamental statistical mechanics theory of percolation as a paradigmatic example of critical phenomena 
\cite{dorogovtsev2008critical,molloy1995critical,newman2001random,newman2009random,kahng009percolation,li2021percolation,lee2018recent,dorogovtsev2022nature}.
We derive the message-passing equations for percolation on factor graphs and hypergraphs and apply them to random hypergraphs and random multiplex hypergraphs. 
In the first model, a 
%%uncorrelated  
random hypergraph is described by two given distributions, namely, a degree distribution for nodes and a cardinality distribution for hyperedges. In the second, more detailed model, a  
random hypergraph is described by a given joint degree distribution, where a degree is a list (i.e., a vector), whose entries are the numbers of hyperedges with each cardinality, adjacent to a node. For these two kinds of random, locally tree-like hypergraphs we obtain the criterion of the presence of a percolation cluster (giant connected component) and the relative size of this cluster.   
We show that, in contrast to ordinary networks, the node and hyperedge percolation problems for hypergraphs strongly differ from each other.  However we highlight that the percolation 
threshold 
for node percolation on hypergraph 
coincides with the percolation threshold of hyperedge percolation on the factor graphs provided one chooses the probability of removing hyperedges of a given cardinality in a suitable way. 

The paper is organized as follows. 
In Sec.~\ref{s0} we define the mapping between hypergraphs and factor graphs.
In Secs.~\ref{s1} and \ref{s2} we remind the key equations for message passing on factor graphs and the relevant analytical results on random factor graphs and random multiplex factor graphs. 
In Sec.~\ref{s3} we derive the equations of the message-passing algorithm for percolation on hypergraphs. 
In Sec.~\ref{s4} we apply these equations to the model of random  hypergraphs and random multiplex hypergraphs with given distributions of node degrees and hyperedge cardinalities and obtain the relative size of the giant connected component and the criterion of its existence.  
In Sec.~\ref{s6} we discuss our results and indicate other problems where results for hypergraphs and their factor graphs may be distinct.

%%%%%%%%%%%%%%%
%%%%%%%%%%%%%%%
%%%%%%%%%%%%%%%

\section{Hypergraphs and their mapping to factor graphs}
\label{s0}

A hypergraph $H=(V,E_H)$ is formed by a set of nodes $V$ and a set of hyperedges $E_H$ where each hyperedge $\alpha$ of cardinality $m_{\alpha}$ indicates a set of $m_{\alpha}$ nodes 
\bea
\alpha=[v_1,v_2,\ldots v_{m_{\alpha}}].
\eea
We indicate with $N=|V|$ the cardinality of the node set and with $M=|E_H|$ the cardinality of the hyperedge set.

\begin{figure}
\begin{center}
\includegraphics[width=\columnwidth]{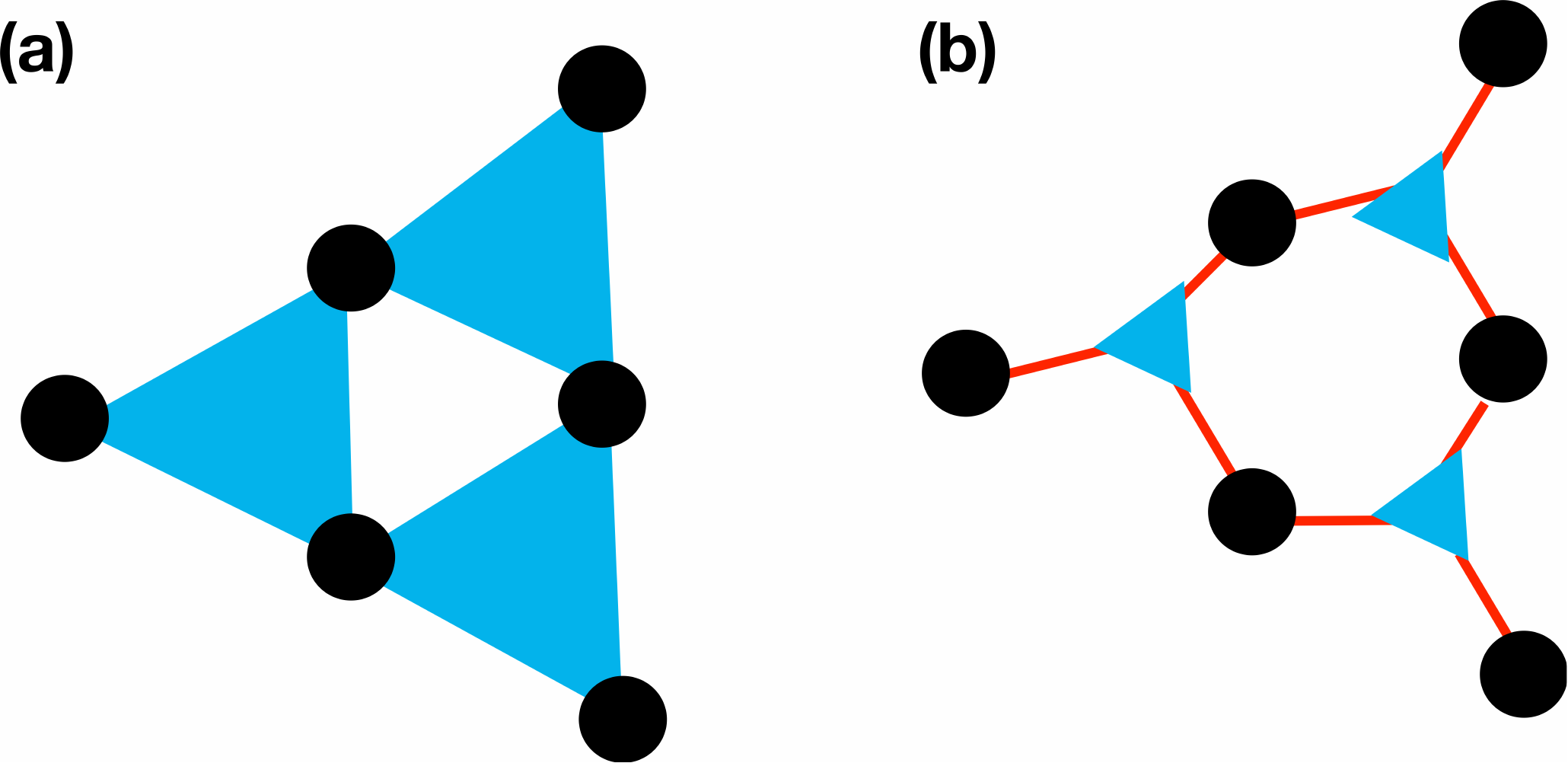}
\end{center}
\caption{Schematic representation of a hypergraph (panel a) and its corresponding factor graph (panel b). The factor graph is a bipartite network of connections between nodes (filled circles) representing  the nodes of the hypergraph and factor nodes (filled triangles) representing  the hyperedges of the hypergraph with a factor node connected to a node in the factor graph if and only if 
this node 
is incident to the corresponding hyperedge in the hypergraph.
} 
\label{fig:fh_1}
\end{figure}

A hypergraph $H=(V,E_H)$ can always be represented as a factor graph (see Fig.~\ref{fig:fh_1}).
The factor graph is a bipartite network 
 $G=(V,U,E)$ formed by a set of nodes $V$, a set of factor nodes $U$, and a set of edges each one linking one node  and one factor node. 
 The factor graph  $G=(V,U,E)$ represents the hypergraph $H=(V,E_H)$ when the set of nodes $V$ of the factor graph coincides with the set of nodes of the hypergraph and when each factor node $\alpha\in U$ represents the hyperedge $\alpha\in E_H$. Therefore we have $N=|V|$ and $M=|U|$. In this setting the factor graph $G$ uniquely represents the hypergraph $H$ if each  node $i\in V$ is linked to the factor node $\alpha\in U$ if and only if the node $i\in V$ belongs to the hyperedge $\alpha\in E_H$ in the corresponding hypergraph. It follows that in this mapping between hypergraphs and factor graphs, a hyperedge $\alpha$ of cardinality $m$ maps into a factor node of degree $m$.

Percolation processes characterize the size of the giant connected component when nodes or factor nodes/hyperedges are randomly removed.

In this work our goal is to highlight the similarity and differences between percolation on  hypergraphs and on their corresponding factor graphs. These two problems coincide when the hyperedges of the hypergraph and hence the corresponding factor nodes of the factor graph are randomly removed but differ when nodes are randomly removed.

%%%%%%%%%%%%%%%%%%%%%%

%%%%%%%%%%%%%%%%%%%%%%

%%%%%%%%%%%%%%%%%%%%%%

\begin{figure}
\begin{center}
\includegraphics[width=\columnwidth]{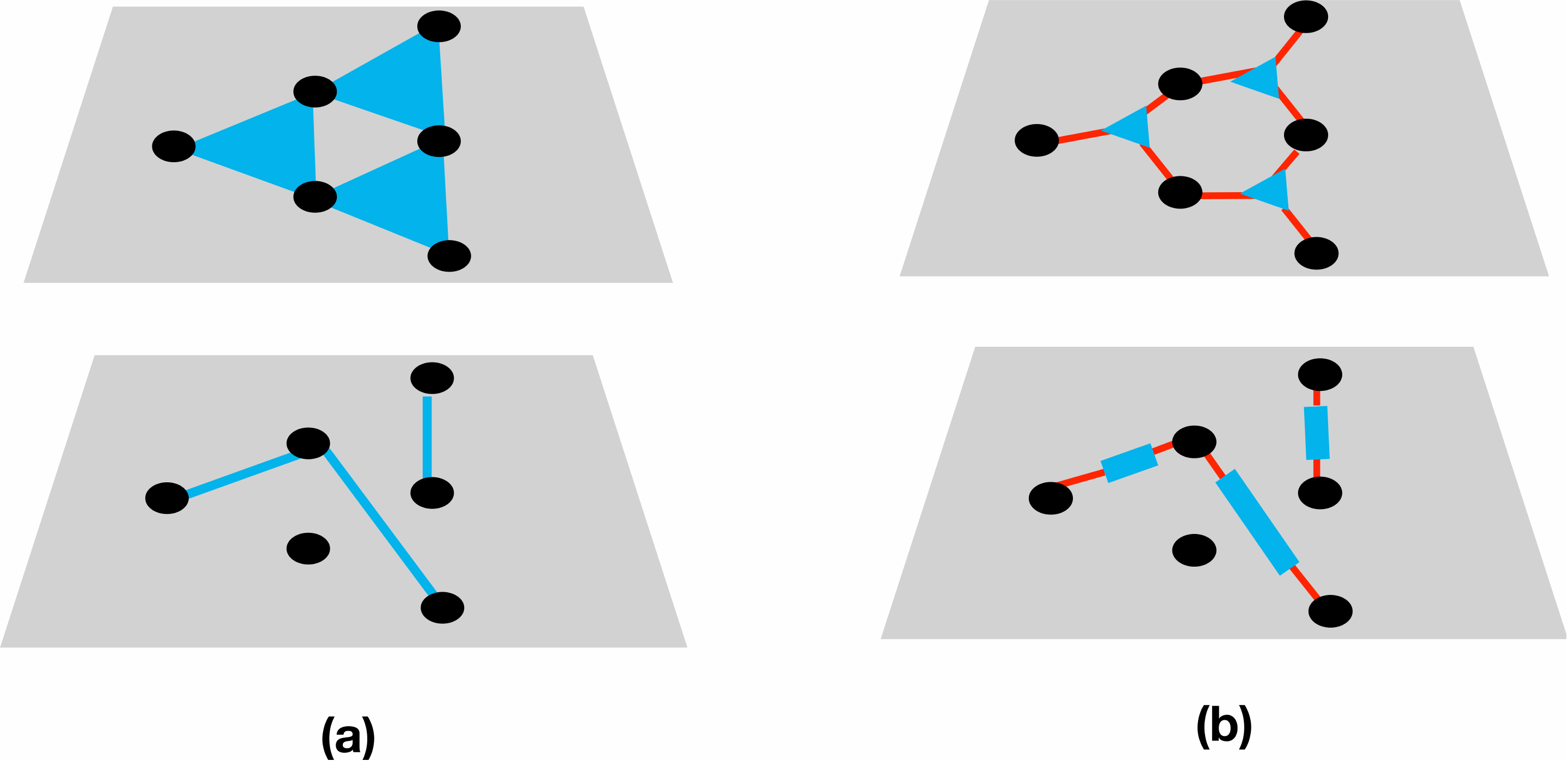}
\end{center}
\caption{A multiplex hypergraph (panel a) \cite{sun2021higher,ferraz2021phase} is a non-uniform (i.e. containing hyperedges with different cardinalities) hypergraph respresented by a multiplex network in which each layer $m$ only accounts for the hyperedge of cadinality $m$ in the hypergraph. In panel a the multiplex hypergraph has two layers $m=2$ and $m=3$. A multiplex hypergraph can be mapped to a multiplex factor graph (panel b) in which each layer is formed by the factor graph corresponding to the hypergraph in corresponding layer of the multiplex hypergraph. When each layer of the hypergraph is drawn from a random hypergraph ensemble, the multiplex hypergraph in general will be correlated. Similarly the corresponding multiplex factor graph will also be correlated.
} 
\label{fig:fh_2}
\end{figure}

%%%%%%%%%%%%%%%%%%%%%%

\section{Message passing for factor graph percolation}
\label{s1}

Factor graph percolation monitors the size of the giant connected component, i.e. the fraction of nodes and/or the fraction of factor nodes in the giant connected component of the factor graph, when either nodes or factor nodes are randomly removed.
Let us define the message-passing algorithm which implements percolation on the factor graph on locally tree-like factor graphs. 
 {In the locally tree-like bipartite networks, finite cycles virtually vanish as network sizes approach infinity, and almost all cycles in these networks are infinite \cite{newman2001random}. In other words, the probability that a node of any kind belongs to a finite cycle vanishes in these infinite networks. We emphasize that this notion doesn't differ from that for ordinary networks and hypergraphs. Importantly, the factor graph of a hypergraph is locally tree-like if and only if the hypergraph is locally tree-like. The message-passing algorithms are exact for all locally tree-like networks (including ordinary one-partite networks, multipartite networks, and hypergraphs), and they are approximate for networks having a significant fraction of finite cycles, which is typical for real-world networks.} 
The message-passing algorithm is formulated  on the factor graph $G=(V,U,E)$  and consists in updating recursively messages sent by nodes to factor nodes and messages sent by factor nodes to nodes. These messages are then used to predict the probability that each node and each factor node belong to the giant component.

We distinguish between two types of message-passing algorithms that implement factor graph percolation depending of the type of information that is available. In particular, the first algorithm assumes that we exactly know the initial damage configuration, namely the set of the removed nodes and the removed factor nodes (importantly, this array does not include the additional damage induced by these removals), while the second algorithm assumes that the initial damage configuration is unknown and we have only access to the probabilities that nodes and factor nodes are left intact by the initial damage. 

To this end, in order to define the message-passing algorithm implementing factor graph  percolation we first assume that we know whether each node $i$ is  initially damaged, $x_i=0$, or it is intact, $x_i=1$, and whether each factor node $\alpha$ is initially damaged, $y_{\alpha}=0$, or it is intact, $y_{\alpha}=1$. 

Let us indicate with $N(i)$ the set of factor nodes $\alpha$ that are the neighbors of node $i$ and  with $N(\alpha)$  the set of nodes that are the neighbors of factor node $\alpha$.

The message-passing algorithm consists in updating the messages $\hat{w}_{i\to\alpha}$ and $\hat{v}_{\alpha\to i}$ going from node $i$ to factor node $\alpha$ and from factor node $\alpha$ to node $i$ respectively.
The message $\hat{w}_{i\to\alpha}$ is equal to one, i.e. $\hat{w}_{i\to\alpha}=1$ if 
\begin{itemize}

\item 
node $i$ is not initially damaged, i.e. $x_i=1$; 

\item 
node $i$ receives at least one positive message $\hat{v}_{\beta\to i}=1$ from at least one of its neighbor factor nodes $\beta$ different from $\alpha$, indicating that $\beta$ is connected to the giant connected component of the factor graph.

\end{itemize}
In all the other cases $\hat{w}_{i\to\alpha}=0$. 
\\
The message $\hat{v}_{\alpha\to i}$ is set to one, i.e. $\hat{v}_{\alpha\to i}=1$ if 
\begin{itemize}

\item 
the factor node $\alpha$ is not initially damaged, i.e. $y_{\alpha}=1$; 

\item 
at least one of the neighbor nodes  $j$ of the factor node $\alpha$ that is different from node $i$ is connected to the giant component. This event occurs if the factor node $\alpha$ receives at least one positive message $\hat{w}_{j\to \alpha}=1$ from one of its neighbor nodes $j$ different from $i$. 

\end{itemize}
In all the other cases $\hat{v}_{\alpha\to i}=0$. 
\\ 
Consequently these messages are updated according to the following rules:
\bea 
\hat{w}_{i\rightarrow \alpha} &=& x_i \left[ 1 - \prod_{\beta\in N(i)\setminus \alpha}(1 - \hat{v}_{\beta\rightarrow i}) \right],\nonumber 
\\
%%\label{Smp_2aa}
\hat{v}_{\alpha\rightarrow i} &=& y_{\alpha}\left[ 1 - \prod_{j\in N(\alpha)\setminus i}(1 - \hat{w}_{j\rightarrow \alpha}) \right]
.
\label{Smp_2aa}
%%\label{Smp_1aa}
\eea

The indicator function $\hat{r}_i$  that node $i$ is in the giant connected component is $1$ ($\hat{r}_i=1$) if and only if 
\begin{itemize}

\item node $i$  is not initially damaged, i.e. $x_i=1$;

\item node $i$ receives at least one positive message $\hat{v}_{\alpha\to i}=1$ from at least one of its neighbor factor nodes.

\end{itemize}
In all the other cases $\hat{r}_i=0$. 

The indicator function $\hat{s}_\alpha$   that the factor node $\alpha$ is in the giant component is $1$ ($\hat{s}_\alpha=1$) if and only if
\begin{itemize}

\item the factor node is not initially damaged, i.e. $y_{\alpha}=1$;

\item at least one of the neighbor nodes  $i$ of the factor node  $\alpha$  is connected to the giant component. This event occurs if the factor node $\alpha$ receives at least one positive message $\hat{w}_{i\to \alpha}=1$ from one of its neighbor nodes.

\end{itemize}
In all the other cases $\hat{s}_\alpha=0.$\\
We have therefore that $\hat{r}_i$ and $\hat{s}_\alpha$ are defined in terms of the messages as
\bea 
\hat{r}_{i} &=&  x_i\left[ 1 - \prod_{\alpha\in N(i)}(1 - \hat{v}_{\alpha\rightarrow i}) \right],\nonumber \\
\hat{s}_{\alpha} &=& y_{\alpha}\left[ 1 - \prod_{i\in N(\alpha)}(1 - \hat{w}_{i\rightarrow \alpha}) \right].
\label{Smp_1b}
\eea
In a number of situations, however, although we 
might 
have access to the real hypergraph structure, we might not have full knowledge of the initial damage of the nodes.
 {In particular, we might only know that nodes and factor nodes of degree $m$ are initially intact (i.e, not damaged) independently at random with probability $p_N$ and $p_H^{[m]}$, respectively, and that the probability of the initial 
damage 
configuration 
%%of intact nodes and factor nodes 
$\{x_i\}=\{x_1,x_2,\ldots, x_N\}$, $\{y_{\alpha}\}=\{u_{\alpha_1},x_{\alpha_2},\ldots, x_M\}$ 
%%, respectively,  
is} 
\bea
P(\{x_i\},\{y_{\alpha}\})&=&\prod_{i=1}^Np_N^{x_i}(1-p_N)^{1-x_i}
\nonumber 
\\
&&
\!\!\!\!\!\!
\times \prod_{\alpha=1}^M\Big(p_H^{[m_{\alpha}]}\Big)^{y_\alpha}\Big(1-p_H^{[m_{\alpha}]}\Big)^{1-y_\alpha}.
\label{Pxy}
\eea
Consequently in this scenario the message-passing algorithm  needs to predict the fraction of nodes and factor in the giant component using only the probabilities $p_N$ and $p_H^{[m]}$.

In this case we  should consider an alternative message-passing algorithm in which the messages sent from nodes to factor nodes are indicated by ${w}_{i\to\alpha}$ and the messages sent from factor nodes to nodes are indicated with $v_{\alpha\to i}$. The message ${w}_{i\to\alpha}$ is obtained by averaging $\hat{w}_{i\to\alpha}$ over the probability distribution $P(\{x_i\},\{y_{\alpha}\})$ and similarly $v_{\alpha\to i}$ is obtained by averaging $\hat{v}_{\alpha\to i}$ over the probability distribution $P(\{x_i\},\{y_{\alpha}\})$.
In this way we obtain the following message-passing algorithm 
\bea 
w_{i\rightarrow \alpha} &=&   p_N\left[ 1 - \prod_{\beta\in N(i)\setminus \alpha}(1 - v_{\beta\rightarrow i}) \right], 
\nonumber 
\\
v_{\alpha\rightarrow i} &=& p_H^{[m_{\alpha}]}\left[ 1 - \prod_{j\in N(\alpha)\setminus i}(1 - w_{j\rightarrow \alpha}) \right]
.
\label{Smp_1b}
\eea

In this framework the probability  $r_i$ that node $i$ is in the giant component and the probability $s_{\alpha}$ that factor node $\alpha$  is in the giant component are given by  
\bea 
r_{i} &=& p_N \left[ 1 - \prod_{\alpha\in N(i)}(1 - v_{\alpha \rightarrow i}) \right],
\nonumber 
\\ 
s_{\alpha} &=& p_H^{[m]}\left[ 1 - \prod_{i\in N(\alpha)}(1 - w_{i\rightarrow \alpha}) \right]
,
\label{Rmp_1b}
\eea
where these probabilities can be 
%%simply 
obtained by averaging $\hat{r}_i$ and $\hat{s}_{\alpha}$ over the probability distribution $P(\{x_i\},\{y_{\alpha}\})$.
%\end{document}
Node percolation on a factor graph implies characterizing  the fraction of  nodes $R$ and the fraction $S$ of factor nodes in the giant component,  
\bea
R=\frac{1}{N}\sum_{i=1}^Nr_i
,
\nonumber 
\\
S=\frac{1}{M}\sum_{\alpha=1}^Ms_{\alpha}.
\label{S_marginals}
\eea
as a function of $p_N$ and $p_H^{[m]}$. 
In particular, putting $p_H^{[m]}=1$ for all values of $m$ one can characterize node percolation on the factor graph as a function of $p_N$ while by putting $p_N=1$ and $p_H^{[m]}=p_H$ for all $m$ we can characterize factor node percolation on the factor graph.
Both transitions are continuous and occur at a percolation threshold that can be determined by linearising the message-passing algorithm. Indeed for  $w_{i\to\alpha}\ll1$ and $v_{\alpha\to i}\ll1$, Eqs.~(\ref{Smp_1b}) can be linearized to obtain 
\bea
w_{i\to\alpha}&=&p_N\sum_{\beta\in N(i)\setminus \alpha} v_{\beta\to i}
,
\nonumber 
\\
v_{\alpha\to i}&=&p_H^{[m_{\alpha}]}\sum_{j\in N(\alpha)\setminus i} w_{j\to \alpha}
.
\eea
This linearized equation admits a non-zero solution if and only if
\bea
\Lambda(p_N,{\bf p}_H)>1
,
\eea
where $\Lambda(p_N,{\bf p}_H)$ with ${\bf p}_H=(p_H^{[2]},p_H^{[3]},\ldots)$ is the maximum eigenvalue of the factor graph non-backtracking matrix $\mathcal{A}$ of block structure 
\bea
\mathcal{A}=\left(\begin{array}{ll}0 &\mathcal{B}^{NH}\\
\mathcal{B}^{HN}& 0\end{array}\right)
\label{1090}
\eea
with the non-backtracking matrices
$\mathcal{B}^{NH}$ and  $\mathcal{B}^{HN}$ having elements given by 
\bea
\mathcal{B}^{NH}_{(i\to \alpha);(\beta\to j)}&=&p_N\delta_{j,i}(1-\delta_{\beta,\alpha})
,
\nonumber 
\\
\mathcal{B}^{HN}_{(\alpha\to i);( j\to \beta)}&=&p_H^{[m_{\alpha}]}\delta_{\alpha,\beta}(1-\delta_{i,j})
,
\label{1100}
\eea
where $\delta_{x,y}$ indicates the Kronecker delta.
Therefore the transition occurs for 
\bea
 \Lambda(p_N,{\bf p}_H)=1.
\eea
For a uniform damage, i.e. $p_H^{[m_{\alpha}]} = p_H$ independent of $m_{\alpha}$, two percolation thresholds for an arbitrary locally tree-like factor graph coincide, $p_{cH}^\text{(factor graph)} = p_{cN}^\text{(factor graph)}$, which can be obtained from the block matrix $\mathcal{A}$, Eqs.~\eq{1090} and \eq{1100}.

%%%%%%%%%%%%%%%%%%%%%%
%%%%%%%%%%%%%%%%%%%%%%
%%%%%%%%%%%%%%%%%%%%%%
%%%%%%%%%%%%%%%%%%%%%%

\section{Analytical solution of factor graph percolation}
\label{s2}

\subsection{Percolation on a random factor graph}

%The message passing algorithms can be used to predict the size of the giant component for arbitrary real world factor graphs as long as their corresponding factor graph is locally tree-like. 
 The message-passing algorithms from Sec.~\ref{s1} enable us to obtain the giant connected component for an arbitrary factor graph as long as its structure is close to a locally tree-like one. 
However in a number of cases we do not have access to the full topology of the factor graph and we only know its structural statistical properties.
In this case we can get analytical predictions for factor graph percolation as long as we assume that the factor graph can be modelled by a random bipartite network. In particular we assume that the factor graph is a random sparse bipartite network  where the nodes have a degree distribution $P(q)$ and the factor nodes have a degree distribution $Q(m)$, where $P(q)$ and $Q(m)$ are arbitrary, provided the total number of links 
incident to the nodes is equal to the total number of links of the factor nodes, i.e. $N \avg{q} = M \avg{m}$.
The factor graph $G$ is uniquely determined by the $N\times M$ incidence matrix ${\bf a}$ of elements $a_{i\alpha}$, where $a_{i\alpha}=1$ if node $i$  is connected to factor node $\alpha$ and  otherwise $a_{i\alpha}=0$.
We assume that the factor graph $G$ is drawn from the distribution 
\bea
P(G)=\prod_{i,\alpha} p_{i\alpha}^{a_{i\alpha}} (1 - p_{i\alpha})^{1-a_{i,\alpha}}
,
\label{PG}
\eea
with $p_{i\alpha}$ indicating the probability that in an uncorrelated factor graph node $i$ is connected to factor node $\alpha$, 
\bea
p_{i\alpha}=\frac{q_im_{\alpha}}{\avg{m}M}
.
\eea
Here the degree sequences $\{q_i\}$ and $\{m_{\alpha}\}$ of the nodes and of the factors nodes are drawn from the  distribution $P(q)$ and $Q(m)$. 
In the limit $N\to \infty$ and $M\to \infty$ with  $M/N=O(1)$, standard percolation describes a critical phenomenon 
that can be studied with statistical mechanics approaches. 

In particular, factor graph percolation can be captured by two nonlinear equations: (i) for the probability  $W$, that if we follow a random edge of a factor node, then we reach a node in the giant component, and (ii) for the probability $V$, that by following a random edge of a  node we reach a factor node  in the giant component.  The probability $W$ can be obtained by averaging the messages $w_{\alpha\to i}$ over $P(G)$ and, similarly, the probability $V$ can be obtained by averaging the messages $v_{i\to\alpha}$ over the probability $P(G)$. 
In this way, starting from the message-passing algorithm Eqs.~(\ref{Smp_1b}), it is straightforward to derive the following equations for $W$ and $V$:  
\bea 
W &=& p_N \sum_m \frac{qP(q)}{\avg{q}} \left[ 1 - (1 - V)^{q-1} \right], 
\nonumber 
\\
V &=&  \sum_m p_H^{[m]}\frac{mQ(m)}{\avg{m}} \left[ 1 - (1 - W)^{m-1} \right].
\label{20}
\eea
The fractions $R$ and $S$ of the, respectively, nodes and factor nodes in the giant connected component are expressed in terms of the probabilities $W$ and $V$,    
\bea 
R &=& p_N \sum_q P(q) \left[ 1 - (1 - V)^q \right],
\nonumber 
\\ 
S &=&  \sum_m p_H^{[m]} Q(m) \left[ 1 - (1 - W)^m \right]
,
\label{40}
\eea
where these equations are obtained by averaging Eqs.~(\ref{Rmp_1b}) and (\ref{S_marginals}) over $P(G)$.
Assuming that both the degree distributions $P(q)$ and $Q(m)$ have finite second moments, one can linearize the right-hand sides of Eqs.~\eq{20}, which leads to the following criterion of the existence of the giant connected component in this network:
\be
 p_N \frac{\avg{q(q-1)}}{\avg{q}} \frac{\Avg{p_H^{[m]}m(m-1)}}{\avg{m}} > 1
.
\label{50}
\ee
In particular if $p_H^{[m]}=p_H$ do not depend on  $m$, then we obtain 
\bea
 p_N p_H\frac{\avg{q(q-1)}}{\avg{q}} \frac{\Avg{m(m-1)}}{\avg{m}} > 1
.
\label{50b}
\eea 
One can check that the phase transition occurring when the left-hand side of Eq.~\eq{50} equals $1$ is continuous. 
Setting $p_H=1$ in Eq.~(\ref{50b}) we obtain the node percolation threshold $p_{cN}^\text{(factor graph)}$, and setting $p_N=1$ we obtain the factor node percolation threshold $p_{cH}^\text{(factor graph)}$. These two thresholds coincide similarly to ordinary uncorrelated networks.  
Hence we have $p_{cN}^\text{(factor graph)} = p_{cH}^\text{(factor graph)} = p^{\star}$ 
with $p^{\star}$ satisfying
\bea
 p^{\star}\frac{\avg{q(q-1)}}{\avg{q}} \frac{\Avg{m(m-1)}}{\avg{m}} = 1
.
\label{50c}
\eea
This equation and Eq.~\eq{50} can be compared with the Molloy--Reed criterion $p \avg{q(q-1)}/\avg{q} > 1$ for ordinary networks \cite{molloy1995critical}, which is valid both for the node 
and edge percolation problems when $\avg{q^2}$ is finite.

%%%%%%%%%%%%%%%%%%%%%%
%%%%%%%%%%%%%%%%%%%%%%

\subsection{Generalized degree distribution}

The random bipartite network is not the only possible ensemble in which factor graph percolation can be solved analytically.
Indeed, it is possible for any non-uniform hypergraph (and corresponding factor graph) to construct a multiplex hypergraph \cite{sun2021higher,ferraz2021phase} (see Fig.~\ref{fig:fh_2}) encoding for its structure, where each layer is formed by a random hypergraph 
having hyperdges of fixed cardinality. These leads to a correlated hypergraph (and corresponding factor graph) structure in which each node  $i$ is associated a vector of degrees ${\bf q}(i)=\{q_2(i),q_3(i)\ldots q_{m}(i)\ldots \}$ where $q_m(i)$ indicates the number of hyperedges of cardinality $m$ to which node $i$ belongs and each hyperedge $\alpha$ of cardinality $m_{\alpha}=m$ describes one higher-order interaction in the layer $m$.
The corresponding factor graph ensemble is a multiplex ensemble of factor graphs $\vec{G}=(G^{[2]},G^{[3]},\ldots G^{[m]}\ldots)$, such that in each network $G^{[m]}$ in layer $m$ the generic node $i$ has degree $q_m(i)$ and all the factor nodes have constant degree $m$.
Let us indicate with the matrix ${\bf a}^{(m)}$ the incidence matrix 
of the factor graph in the $m$ layer of the multiplex factor graph.
The probability of this multiplex factor graph $\vec{G}$ is given by 
\bea
\!\!\!\!\!\!\!\!\!\!\!\!\! 
P(\vec{G})=\prod_{m\geq 2}\prod_{\alpha\in \mathcal{K}(m)}\prod_i[p_{i\alpha}^{(m)}]^{a_{i\alpha}^{(m)}}\left(1-p_{i\alpha}^{(m)}\right)^{1-a_{i\alpha}^{(m)}}
\!,
\label{MultiplexG}
\eea
where $\mathcal{K}(m)$ indicates the set of hyperedges of cardinality $m$ and where  
the  probability $p_{i\alpha}^{(m)}$ of observing a link between node $i$ and factor node $\alpha$ in layer $m$ is given by 
\bea
p_{i\alpha}^{(m)}=\frac{q_{m}(i)}{\Avg{q_m}N}
,
\eea
where $\Avg{q_m}N=mM_{m}$. 
In this ensemble, factor graph percolation is captured by  the probability $W_m$ that by following a random edge of a factor node in layer $m$ we reach a node in the giant component and by the probability $V_m$ that by following a random edge of a node in layer $m$ we reach a factor node in the giant component. These probabilities obey the equations:
\bea
W_m&=&p_{N}\sum_{{\bf q}}\frac{q_m P({\bf q})}{\avg{q_m}}\left[1-\prod_{m'=2}^{M}(1-V_{m'})^{q_{m'}-\delta_{m,m'}}\right]
,
\nonumber 
\\
V_{m}&=&p^{[m]}_{H}\left[1-(1-W_m)^{m-1}\right]
,
\eea
which can be obtained directly by averaging the message-passing algorithm Eqs.~(\ref{Smp_1b}) over the probability $P(\vec{G})$ given by Eq.~(\ref{MultiplexG}). 
Here $q_m P({\bf q})/\avg{q_m}$ is the degree distribution of an end node of an edge in layer $m$.  
The probabilities $R$  and $S$ that a node and a factor node are in the giant component of the multiplex factor graph are given by 
\bea
R&=&p_N\sum_{{\bf q}}P({\bf q})\left[1-\prod_{m'=2}^{M}(1-V_{m'})^{q_{m'}}\right]
, 
\nonumber 
\\
S&=&\sum_{m}p_H^{[m]}Q(m)
%%p_H^{[m]}
\left[1-(1-W_m)^{m}\right
].
\eea
These equations describe a continuous second order phase transition whose critical point is determined by the condition 
\bea
\Lambda=1,
\eea
where $\Lambda$ is the maximum eigenvalue of the matrix ${\bf G}$ of elements
\bea
G_{m,m'}=p_{N}p^{[m]}_{H}(m-1)\frac{\avg{q_{m}(q_{m}-\delta_{m',m})}}{\avg{q_{m}}}
.
\eea
This network has a giant connected component when $\Lambda>1$.

%%%%%%%%%%%%%%%
%%%%%%%%%%%%%%%
%%%%%%%%%%%%%%%

\section{Message-passing algorithm for hypergraph percolation}
\label{s3}

While factor node percolation on a factor graph fully accounts for hyperedge percolation on the corresponding hypergraph, node percolation on a factor graph and on an hypergraph are distinct. In node percolation on a factor graph, a factor node (hyperedge) is still able to connect its nodes also if one or more  of its nodes are initially damaged, provided that at least one of its nodes is connected to the giant component. 
However in node percolation on hypergraphs, a hyperedge is only able to connect a node to the giant component if none of its other nodes are initially damaged.
In other words, the initial damage of a single node of an hyperedge deactivates the entire hyperedge.

Here and in the following we formulate the message-passing algorithms that fully account for this difference.
Interestingly, the message passing for hypergraph percolation is still conveniently  defined on the factor graph representing the hypergraph,  although it  fully accounts for the differences between factor graph and  hypergraph percolation.
We assume that we have full information about the hypergraph structure  completely encoded in the corresponding factor graph wiring.

The message-passing algorithm for hypergraph percolation can be implemented for any real-world hypergraph topology as long as the factor graph corresponding to the hypergraph is locally tree-like.

As we did for the message-passing algorithms for factor graph percolation, we will first formulate the message-passing algorithm under the assumption that  we will have full knowledge of the initial damage.  
Subsequently, we will formulate the message-passing algorithm that can be applied to the case in which we have only access to the probability that nodes and hyperedges are randomly removed.
 
To start with, let us formulate the message-passing algorithm able to predict hypergraph percolation when we have full information about the entity of the initial damage.
To this end, let us  assume that we know whether each node $i$ is initially damaged $x_i=0$ or intact $x_i=1$ and whether each hyperedge $\alpha$ is initially damaged $y_{\alpha}=0$ or intact $y_{\alpha}=1$. 
Note that the variable $y_{\alpha}$ does not take into account the damage induced by the removal of nodes. 
The message-passing algorithm implementing hypergraph percolation consists in updating the messages $\hat\omega_{i\to\alpha}$  and $\hat{v}_{\alpha\to i}$ going from node $i$ to factor node $\alpha$ and from factor node $\alpha$ to node $i$.
The message $\hat\omega_{i\to\alpha}$ is equal to $1$, i.e. $\hat\omega_{i\to\alpha}=1$,  if 
\begin{itemize}

\item 
node $i$  is not initially damaged, i.e. $x_i=1$;  

\item 
node $i$ receives at least one positive message $\hat{v}_{\beta\to i}=1$ from at least one of its neighbor factor nodes $\beta$ different from $\alpha$, indicating that it is connected to the giant component of the hypergraph. 

\end{itemize}
In all the other cases $\hat\omega_{i\to\alpha}=0$. 
\\
The message $\hat{v}_{\alpha\to i}$ is equal to $1$, i.e. $\hat{v}_{\alpha\to i}=1$ if 
\begin{itemize}

\item 
the hyperedge is not initially damaged, i.e. $y_{\alpha}=1$; 

\item 
each of the nodes $j$ that belong to hyperedge $\alpha$ and differ from the node $i$ is intact, i.e. $\prod_{j\in N(\alpha)\setminus i}x_j=1$; 

\item 
at least one of the nodes  $j$ that belong to hyperedge $\alpha$ and differ from the node $i$ is connected to the giant component. This event occurs if the factor node $\alpha$ receives at least one positive message $\hat\omega_{j\to \alpha}=1$  from one of its neighbor nodes different from $i$. 

\end{itemize}
In all the other cases $\hat{v}_{\alpha\to i}=0.$\\
Consequently these messages are updated according to the following rule
\bea 
\!\!\!\!\!\!\!\!\!\!\!\!
\hat{\omega}_{i\rightarrow \alpha} &=& x_i \left[ 1 - \prod_{\beta\in N(i)\setminus \alpha}(1 - \hat{v}_{\beta\rightarrow i}) \right]
, 
\nonumber 
\\
\!\!\!\!\!\!\!\!\!\!\!\! 
\hat{v}_{\alpha\rightarrow i} &=& y_{\alpha}\!\left(\prod_{j\in N(\alpha)\setminus i}x_i\right)\!\!\left[ 1 - \!\prod_{j\in N(\alpha)\setminus i}\!\!(1 - \hat{\omega}_{j\rightarrow \alpha}) \right]
.
\label{mp_1aa}
\eea
These equations can be further simplified by introducing a message  $\hat{w}_{i\to \alpha}$ that indicates the message sent by a  node $i$ to a neighbor factor node $\alpha$,  under the assumption that $x_i=1$, i.e.
\bea
\hat{w}_{i\rightarrow \alpha} &=&  \left[ 1 - \prod_{\beta\in N(i)\setminus \alpha}(1 - \hat{v}_{\beta\rightarrow i}) \right]
,
\eea
which is related to the previously defined message $\hat{\omega}_{i\to\alpha}$ by
\bea
\hat{\omega}_{i\to \alpha}=x_i\hat{w}_{i\to \alpha}
.
\eea
The message $\hat{v}_{\alpha\to i}$ can now be expressed directly in terms of $\hat{w}_{i\to \alpha}$ as 
\bea
\!\!\!\!\!\!
\hat{v}_{\alpha\to i} &=& y_{\alpha}\left(\prod_{j\in N(\alpha)\setminus i}\!x_i\!\right)\!\!\left[ 1 - \!\prod_{j\in N(\alpha)\setminus i}\!(1 - x_j\hat{w}_{j\rightarrow \alpha}) \right]
\nonumber 
\\
&=&y_{\alpha}\left(\prod_{j\in N(\alpha)\setminus i}\!x_i\!\right)\!\!\left[ 1 - \!\prod_{j\in N(\alpha)\setminus i}\!(1 - \hat{w}_{j\rightarrow \alpha}) \right]
,
\label{mp_1aa}
\eea
where in this expression we have used the fact that $\hat{v}_{\alpha\to i}$ is only non-zero if $x_j=1$ for every $j\in N(\alpha)\setminus i$, hence we can safely substitute $\hat{\omega}_{i\to \alpha}$ with $\hat{w}_{i\to \alpha}$.
The indicator function $\hat{r}_i$  that node $i$ is in the giant component is $1$ ($\hat{r}_i=1$)  if and only if 
\begin{itemize}

\item 
node $i$  is not initially damaged, i.e. $x_i=1$, 

\item 
node $i$ receives at least one positive message ($\hat{v}_{\alpha\to i}=1$) from at least one of its neighbor factor nodes.

\end{itemize}
In all the other cases $\hat{r}_i=0$.  
 {The  indicator function $\hat{s}_\alpha$  that the factor node $\alpha$ is in 
the giant component 
equals 
%%is 
$1$ ($\hat{s}_\alpha=1$)  
if and only if}
\begin{itemize}

\item 
the hyperedge $\alpha$ is not initially damaged; 

\item 
each of the nodes $j$ that belong to the hyperedge $\alpha$ is intact, i.e. $\prod_{j\in N(\alpha)}x_j=1$; 

\item 
at least one of the nodes $j$ that belong to the hyperedge $\alpha$ is connected to the giant component. This event occurs if the factor node $\alpha$ receives at least one positive message $\hat{w}_{i\to \alpha}=1$ from one of its neighbor nodes. 

\end{itemize}
We obtain therefore that the indicator functions $\hat{s}_{\alpha}$  and $\hat{r}_i$ are given by 
\bea 
\!\!\!\!\!\!\!\!\!
\hat{s}_{\alpha} &=& y_{\alpha}\left(\prod_{j\in N(\alpha)\setminus i}x_i\right)\left[ 1 - \prod_{i\in N(\alpha)}(1 - \hat{w}_{i\rightarrow \alpha}) \right]
,
\label{hmp_1b}
\\[3pt] 
\!\!\!\!\!\!\!\!\! 
\hat{r}_{i} &=&  x_i\left[ 1 - \prod_{\alpha\in N(i)}(1 - \hat{v}_{\alpha\rightarrow i}) \right]
.
\label{hmp_2b}
\eea 

In a number of situations, although we might have access to the real hypergraph structure, we might not have full knowledge of the initial damage of the nodes.
In particular, we might only know that nodes and hyperedges of cardinality $m$ are damaged independently at random with probability $p_N$ and $p_H^{[m]}$ respectively, which provides the probability distribution $P(\{x_i\},\{y_{\alpha}\})$ given by Eq.~(\ref{Pxy}).

In this scenario we should consider an alternative message-passing algorithm in which the messages sent from nodes to factor nodes are indicated by ${w}_{i\to\alpha}$ and the messages sent from factor nodes to nodes are indicated with $v_{\alpha\to i}$, where ${w}_{i\to\alpha}$ is obtained by averaging $\hat{w}_{i\to\alpha}$ over the probability distribution $P(\{x_i\},\{y_{\alpha}\})$ and, similarly, $v_{\alpha\to i}$ is obtained by averaging $\hat{v}_{\alpha\to i}$ over the probability distribution $P(\{x_i\},\{y_{\alpha}\})$.
In this way we obtain the following message-passing algorithm:  
\bea 
w_{i\rightarrow \alpha} &=&   \left[ 1 - \prod_{\beta\in N(i)\setminus \alpha}(1 - v_{\beta\rightarrow i}) \right]
,
\label{mp_1b}
\\[3pt] 
v_{\alpha\rightarrow i} &=& p_H^{[m]} p_N^{m_{\alpha}-1}\left[ 1 - \prod_{j\in N(\alpha)\setminus i}(1 - w_{j\rightarrow \alpha}) \right]
,
\label{mp_2b}
\eea
where $m_{\alpha}$ indicates the cardinality of hyperedge $\alpha$ (degree of the corresponding factor node in the bipartite network).
In this framework, the probability $r_i$  that node $i$ is in the giant component and the probability $s_{\alpha}$  that factor node $\alpha$ (hyperedge) is in the giant component are given by  
\bea 
r_{i} &=& p_N \left[ 1 - \prod_{\alpha\in N(i)}(1 - v_{\alpha \rightarrow i}) \right]
,
\label{rp_2b}
\\
[3pt] 
s_{\alpha} &=& p_H^{[m]}p_N^{m_{\alpha}}\left[ 1 - \prod_{i\in N(\alpha)}(1 - w_{i\rightarrow \alpha}) \right]
,
\label{sp_1b}
\eea
where these probabilities can be 
%%simply 
obtained by averaging $\hat{r}_i$ and $\hat{s}_{\alpha}$  over the probability distribution $P(\{x_i\},\{y_{\alpha}\})$.
By comparing the message-passing equations for hypergraph percolation and for factor node percolation, we conclude that the hypergraph percolation algorithm reduces to factor node percolation for $p_N=1$, however 
%%for $p_N\neq 1$ 
the two algorithms differ for $p_N\neq 1$.

Both hyperedge and node percolation are continuous and occur at a percolation threshold that can be determined by linearising the message-passing algorithm. Indeed for  $w_{i\to\alpha}\ll1$ and $v_{\alpha\to i}\ll1$, Eqs.~(\ref{mp_2b}) and (\ref{mp_1b}) can be linearized to obtain 
\bea
w_{i\to\alpha}&=&\sum_{\beta\in N(i)\setminus \alpha} v_{\beta\to i}
,
\nonumber 
\\
v_{\alpha\to i}&=&p_H^{[m_{\alpha}]}p_N^{m_{\alpha}-1}\sum_{j\in N(\alpha)\setminus i} w_{j\to \alpha}
. 
\eea
This linearized equation admits a non-zero solution if and only if
\bea
\Lambda(p_N,{\bf p}_H)>1
,
\eea
where $\Lambda(p_N,{\bf p}_H)$ with ${\bf p}_H=(p_H^{[2]},p_H^{[3]},\ldots)$ is the maximum eigenvalue of the factor graph non-backtracking matrix $\mathcal{A}$ of block structure 
\bea
\mathcal{A}=\left(\begin{array}{ll}0 &\mathcal{B}^{NH}
\\
\mathcal{B}^{HN}& 0\end{array}\right)
\eea
with the non-backtracking matrices
$\mathcal{B}^{NH}$ and  $\mathcal{B}^{HN}$ having elements given by 
\bea
\mathcal{B}^{NH}_{(i\to \alpha);(\beta\to j)}&=&\delta_{j,i}(1-\delta_{\beta,\alpha}),
\nonumber 
\\
\mathcal{B}^{HN}_{(\alpha\to i);( j\to \beta)}&=&p_H^{[m_{\alpha}]}p_N^{m_{\alpha}-1}\delta_{\alpha,\beta}(1-\delta_{i,j}) 
,
\label{3800}
\eea
where $\delta_{x,y}$ indicates the Kronecker delta.
Therefore the condition for the transition is  
\bea
 \Lambda(p_N,{\bf p}_H)=1
 .
\eea

Note that that the hyperedge percolation threshold for a hypergraph coincides with the factor node percolation threshold for the corresponding factor graph. 
We observe that node percolation on factor graph and on hypergraph is distinct, and it can be shown that the percolation threshold on a hypergraph is always strictly larger than for node percolation on factor graphs provided the hypergraph is not only formed by  hyperedges of cardinality  $m=2$, i.e. provided the hypergraph is not a graph. 
Indeed the maximum eigenvalue of the non-negative matrix $\mathcal{A}$ obeys the Collatz-Wielandt formula
\bea
\Lambda=\max_{{\bf x}>0}\min_{\gamma} \frac{{(\mathcal{A}{\bf x})}_\gamma}{x_\gamma}
.
\eea
Now, let us compare the entries of the matrices $\mathcal{A}$ of hypergraphs and factor graphs in the case of $p_H^{[m]}=1$, i.e. for the node percolation problems, Eqs.~\eq{3800} and \eq{1100}. Both matrices have non-negative elements. Furthermore for an arbitrary $p_N$, each entry of the matrix $\mathcal{A}$ of a hypergraph cannot be larger than the corresponding entry of the matrix $\mathcal{A}$ of the factor graph. Consequently, for each choice of the vector ${\bf x}>0$ and each value of $\gamma=(i\to \alpha)$ or $\gamma=(\alpha\to i)$ the ratio ${{(\mathcal{A}{\bf x})}_\gamma}/{x_\gamma}$ for the hypergraph node percolation problem is smaller or equal than the ratio for the factor node percolation problem. This implies that for a given value of $p_N$ the largest  eigenvalue  of the non-backtracking matrix for factor graph node percolation is larger or equal than the largest eigenvalue of the non-bracktracking matrix for hypergraph node percolation. Thus this argument confirms that the node percolation threshold on a hypergraph is either equal or exceeds the  threshold for node percolation on the factor graph. A closer look to the eigenvalue problem will reveal that the equality holds only if the hypergraph reduces to a network.

This mathematical results can be also derived by the following physical argument.
Let us   compare the node and hyperedge percolation thresholds of a hypergraph. 
Inspecting the matrix elements in Eq.~\eq{3800}, we see that the node percolation threshold $p_{cN}$ coincides with the hyperedge percolation threshold if the probability that hyperedge $\alpha$ of cardinality $m_\alpha$ is retained with probability $p_H^{[m_\alpha]} = p_{cN}^{m_\alpha-1} \leq p_{cN}$. 
If we remove hyperedges uniformly (independently of their cardinalities), then the resulting hyperedge percolation threshold $p_{cH}$ must be lower than the maximum probability $p_H^{[m_\alpha]} = p_{cN}^{m_\alpha-1}$ for the nonuniform removal of edges, i.e. $p_{cH} < p_{cN}^{m_{\min}-1} \leq p_{cN}$. 
Thus this argument also confirms that node percolation threshold for a tree-like hypergraph exceeds its hyperedge percolation threshold. 

Now we recall that for a locally tree-like graph, the node and factor node percolation thresholds coincide, $p_{cN}^\text{(factor graph)} = p_{cH}^\text{(factor graph)}$.  
Consequently we have as long as the hypergraph does not reduce to a network, 
\be
p_{cN} > p_{cH} = p_{cH}^\text{(factor graph)} = p_{cN}^\text{(factor graph)} 
%%,
\ee
and hence indeed the node percolation threshold of a locally tree-like hypergraph exceeds the factor node percolation threshold for the corresponding factor graph. 

\begin{figure}[t]
\begin{center}
\includegraphics[width=\columnwidth]{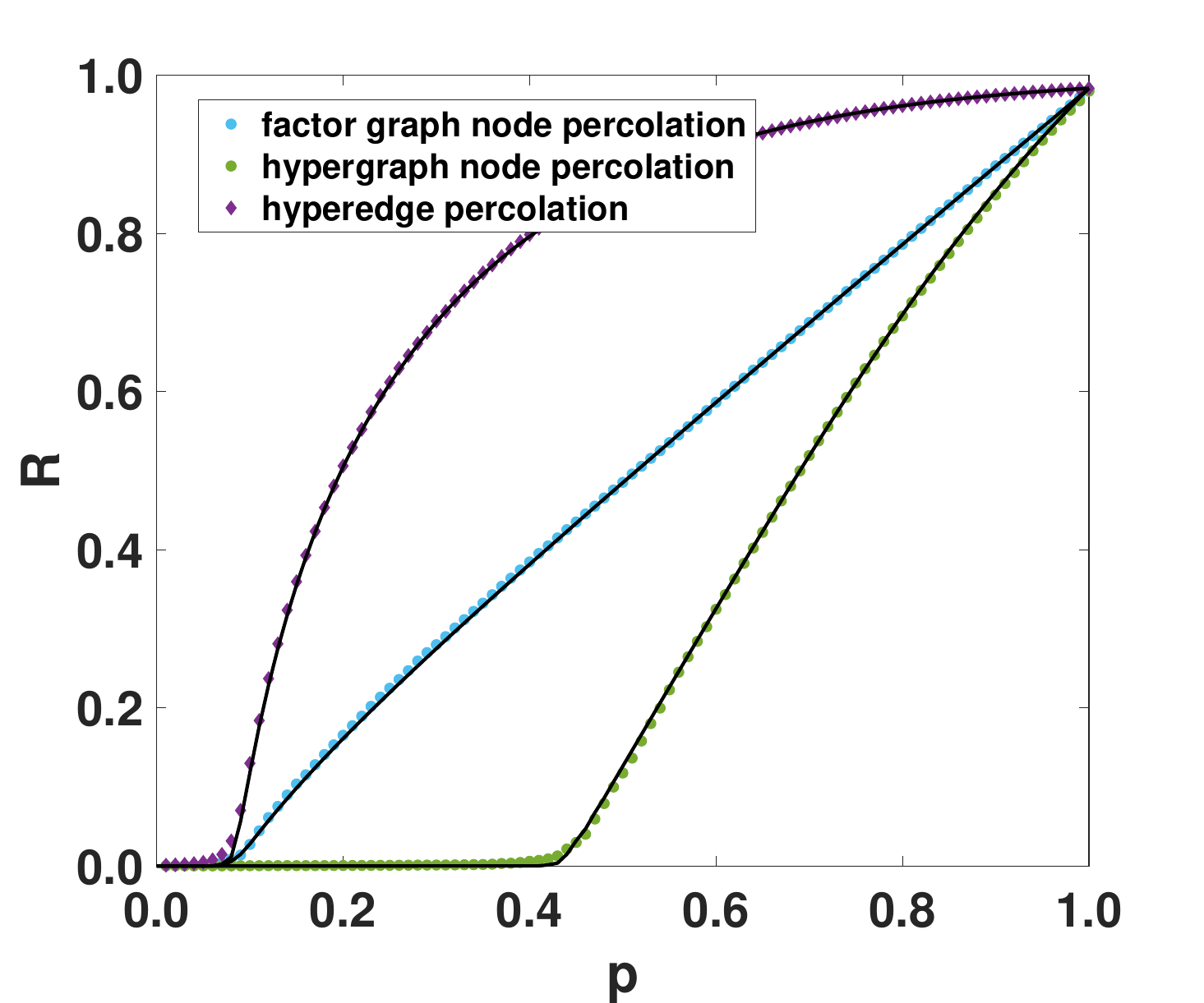}
\end{center}
\caption{Factor graph node 
%%site 
percolation and hypergraph 
node 
%%site 
percolation and hyperedge percolation for a factor graph and the corresponding hypergraph with $N=10^4$ nodes and $M=10^4$ factor nodes (hyperedges)  {having the Poisson degree distribution $P(q)$ with $\avg{q}=4$ for nodes and $Q(m)=\delta(m,4)$ for factor nodes. The symbols indicate results of the Monte Carlo simulations of the percolation processes and the solid lines indicate results obtained with the corresponding message-passing algorithms. 
Hyperedge  percolation on a hypergraph coincides with factor node percolation on the corresponding factor graph. 
The percolation threshold for factor graph percolation (and hyperedge percolation) is $p_c=1/12=0.0833\ldots$, Eq.~\protect\eq{50b}, while for hypergraph node percolation, it is $p_c=12^{-1/3}=0.437\ldots$,} Eq.~\protect\eq{100}. 
} 
\label{f1}
\end{figure}

\begin{figure}[t]
\begin{center}
\includegraphics[width=\columnwidth]{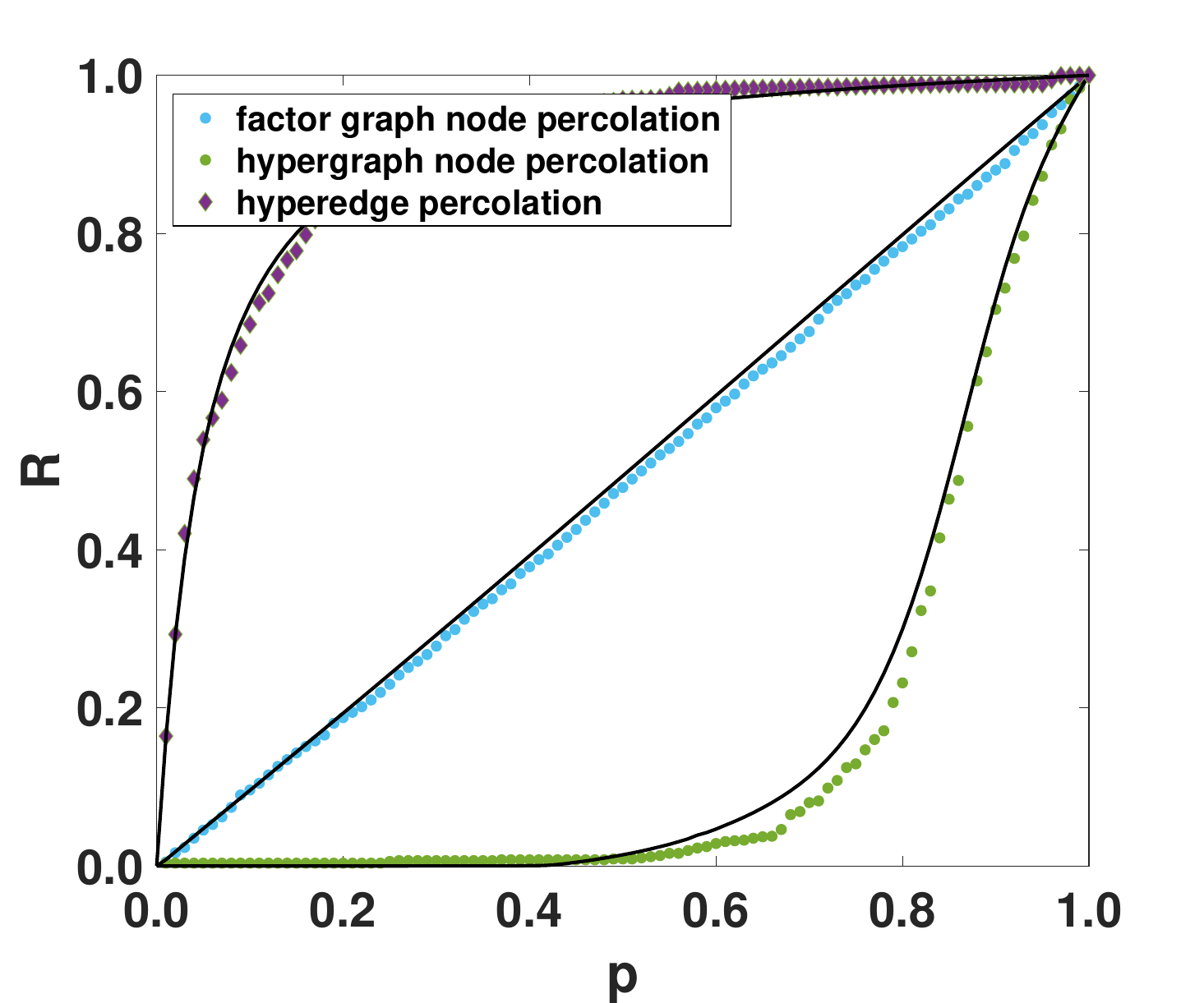}
\end{center}
\caption{Factor graph node percolation and hypergraph node percolation and hyperedge percolation for a factor graph and the corresponding hypergraph  on the US Senate Commitee hypergraph \cite{chodrow2021generative,stewart2008congressional}. The symbols indicate results of the Monte Carlo simulations of the percolation processes and the solid lines indicate results obtained with the corresponding message-passing algorithms.  The dataset has $N=1290$ nodes and $M=341$ hyperedges;  {the average degree of nodes is $\avg{q}=9.2$,} the average cardinality of the hyperedges is $\avg{m}=34.8$, and their maximum cardinality is $m_{\max}=82$.
} 
\label{f2}
\end{figure}

This effect is also evident from our numerical comparison of factor graph percolation and hypergraph percolation on 
%%random hypergraphs 
 {a 
%%regular uniform 
random hypergraph of $10^4$ nodes and $10^4$ hyperedges, having a Poisson degree distribution with $\avg{q}=4$  and all hyperedges of the same cardinality $m=4$,} Fig.~\ref{f1}, and on the real hypergraph of US Senate committees \cite{chodrow2021generative,stewart2008congressional}, see Fig.~\ref{f2}. This is a hypergraph, where nodes are members of the US House of Representatives and hyperedges correspond to commitee memberships. The hypergraph has $1290$ nodes  {with average degree $9.2$} and $341$ hyperedges with the average size of hyperedges $\avg{m}=34.8$ and the maximum size of the hyperedges $m_{\max}=82$. 
 {These two figures compare the results of numerical simulations and of the message-passing algorithm for each of these networks having, notably, rapidly decaying degree and cardinality distributions.} 
The message-passing algorithm  
perfectly describes the indicated difference on the random hypergraph whose corresponding factor graph is tree-like,  
and this algorithm provides still a good prediction of the percolation process also in the case of the real hypergraph which deviates from the pure locally tree-like approximation.
In both cases we find confirmation that the percolation threshold for hypergraph percolation can be significantly higher than the percolation on the corresponding factor graph. 
 {Even for the synthetic random hypergraph with $\avg{q}=4$ and $m=4$, the difference between these thresholds is surprisingly big, namely, $0.44$, vs. $0.08$, and for the real hypergraph of US Senate committees, whose factor graph has a very low percolation threshold, this difference is dramatic, see Fig.~\ref{f2}.}

Interestingly, it turns out that by setting $p_H^{[m]}=\pi^{m-1}$ in the factor graph percolation,  
one finds that the percolation threshold $\pi_c$ and the node percolation threshold of a hypergraph, $p_{cN}$, coincide, i.e. $\pi_c=p_{cN}$. 
That is, the percolation threshold of node percolation on hypergraph coincides with the percolation threshold for factor node percolation on the corresponding factor graph when factor nodes  of degree $m$ are left intact with probability $\pi^{m-1}$. 
Note however that the this mapping does not extent to the fraction of nodes in the giant connected component. 
 
Let us indicate with $R$ the fraction of nodes in the giant component of node percolation on the hypergraph (assuming that $p_N=p\in [0,1]$ and $p_H^{[m]}=1$ for every $m$), 
with $R^\text{(factor graph)}$ the fraction of nodes in the giant connected component for factor graph percolation with $p_H^{[m]}=p^{m-1}$ and $p_N=1$, and let us indicate with $S_m$ and $S^\text{(factor graph)}_m$ the fractions of, respectively, hyperedges of cardinality $m$ and factor nodes of degree $m$ within the giant connected components for these problems.  
Then we have 
\bea
R&=&pR^\text{(factor graph)}
,
\nonumber 
\\[3pt]
S_m&=&p^{m}S^\text{(factor graph)}_m
.
\eea
%%

%%
%%%%%%%%%%%%%%%%%%%%%%%%%%%%%%%%%%%%%%%%%%%%%%%%%%%%%
%%%%%%%%%%%%%%%%%%%%%%%%%%%%%%%%%%%%%%%%%%%%%%%%%%%%%
%%

%%%%%%%%%%%%%%%%%%%%%%%%%%%%%%%%%%%%%%%%%%%%%%%%%%%%%
%%%%%%%%%%%%%%%%%%%%%%%%%%%%%%%%%%%%%%%%%%%%%%%%%%%%%

%%%%%%%%%%%%%%%%%%%%%%%%%%%%%%%%%%%%%%%%%%%%%%%%%
%%%%%%%%%%%%%%%%%%%%%%%%%%%%%%%%%%%%%%%%%%%%%%%%%
%%%%%%%%%%%%%%%%%%%%%%%%%%%%%%%%%%%%%%%%%%%%%%%%%

%%%%%%%%%%%%%%%%%%%%%%%%%%%%%%%%%%%%%%%%%%%%%%%%%
%%%%%%%%%%%%%%%%%%%%%%%%%%%%%%%%%%%%%%%%%%%%%%%%%
%%%%%%%%%%%%%%%%%%%%%%%%%%%%%%%%%%%%%%%%%%%%%%%%%
%%%%%%%%%%%%%%%%%%%%%%%%%%%%%%%%%%%%%%%%%%%%%%%%%

\section{Analytical solution of hypergraph percolation}
\label{s4}

 {\subsection{Percolation for the configuration model of uncorrelated hypergraphs}}

We can apply the same configuration model as in Sec.~\ref{s2} to uncorrelated hypergraphs. 
In this case, a random hypergraph is described by a given node degree distribution $P(q)$ and a given hyperedge cardinality distribution $Q(m)$. 
Similarly to Sec.~\ref{s2}, we introduce the averages $V = \avg{v_{\alpha\to i}}$ and $W = \avg{w_{i\to\alpha}}$ of the messages considered in Sec.~\ref{s3} over the distribution $P(H)$ for a random hypergraph, given by the same Eq.~\eq{PG} as for a random factor graph. 
Averaging Eqs.~\eq{mp_2b} and \eq{mp_1b} over the distribution $P(H)$, we arrive at the self-consistensy equations for $W$ and $V$, 
\bea 
\!\!\!\!\!\! 
W &=&  \sum_q \frac{qP(q)}{\avg{q}} \left[ 1 - (1 - V)^{q-1} \right]
,
\label{70}
\\
[3pt]
\!\!\!\!\!\! 
V &=& \sum_{m=2}^M p_H^{[m]}p_N^{m-1}\frac{mQ(m)}{\avg{m}} \left[ 1 - (1 - W)^{m-1} \right]
.
\label{60}
\eea
Averaging Eqs.~\eq{rp_2b} and \eq{sp_1b} over $P(H)$, we obtain the expressions for the relative size $R = \avg{r_i}$ of the giant connected component in the hypergraph and for the probability $S = \avg{s_\alpha}$ that a hyperedge connects two nodes in the giant connected component, 
\bea 
R &=& p_N \sum_q P(q) \left[ 1 - (1 - V)^q \right]
,
\label{80}
\\[3pt] 
S &=& \sum_m p_H^{[m]}p_N^{m}Q(m) \left[ 1 - (1 - W)^m \right]
.
\label{90}
\eea
Assuming that $p_H^{[m]} = p_H$ is independent of $m$ and that $\avg{q^2}$ is finite and linearizing Eqs.~\eq{60} and \eq{70}, we arrive at the following criterion of the presence of the giant connected component in this problem: 
\be
p_N p_H \frac{\avg{q(q-1)}}{\avg{q}} \sum_m\frac{m(m-1)Q(m)}{\avg{m}} p_N^{m-2} > 1 
.
\label{100}
\ee
Introducing the generating functions $G_P(x) \equiv \sum_q Q(q) x^q$ of the degree distribution $P(q)$ and $G_{1P}(x) \equiv G'_P(x)/G_P(1)$, and $G_Q(x) \equiv \sum_m Q(m) x^m$ of the cardinality distribution $Q(m)$ and $G_{1Q}(x) \equiv G'_Q(x)/G_Q(1)$, we can rewrite this criterion in the compact form:  
\be 
p_N p_H G'_{1P}(1) G'_{1Q}(p_N)  > 1
.
\label{110}
\ee
It is worthwhile to compare the criteria, Eqs.~\eq{50} and \eq{100}. 
The term $p_N^{m-2}$ effectively introduces a cut-off in the cardinality distribution and makes the sum in Eq.~\eq{100} finite even if the first moment of this distribution diverges. 
Since $\sum_m m(m-1)Q(m)p_N^{m-2}/\avg{m}<\avg{m(m-1)}/\avg{m}$ if $p_N<1$ and $Q(m) \neq \delta_{m,2}$, the node percolation threshold for this model of an uncorrelated hypergraph exceeds the hyperedge percolation threshold in contrast to ordinary uncorrelated networks, as we have already showed for general locally tree-like hypergraphs. 

 {Substituting the degree and cardinality distributions $P(Q)$ and $Q(m)$ of the synthetic random hypergraph used for simulations in Fig.~\ref{f1} into Eqs.~\eq{50b} and \eq{100} [or Eq.~\eq{110}], we obtain the exact values of the percolation thresholds for hyperedge percolation (and factor graph percolation) and hypergraph node percolation, $1/12=0.0833\ldots$ and $12^{-1/3}=0.437\ldots$, respectively. The percolation thresholds observed in the simulations agree with these values. 
%%The same equations can be used to estimate roughly the corresponding thresholds of the US Senate Commitee hypergraph used for simulations in Fig.~\ref{f2}. Since this hypergraph has rapidly decaying degree and cardinality distributions
}

Finally, let us consider Eqs.~\eq{60}--\eq{110} in the special case of $Q(m) = \delta_{m,2}$, corresponding to an ordinary networks, where each edge has cardinality $2$.  
In this case, Eqs.~\eq{100} and \eq{110} take the form of the standard Molloy--Reed criterion \cite{molloy1995critical} (node percolation problem). 
Equation~\eq{60} gives $V=pW$, and 
the expression for the relative size of the giant connected component $R$, Eq.~\eq{80}, also turn out to be standard for uncorrelated networks.  
Thus in this case  
our equations properly reduce to the known results for node percolation on uncorrelated networks.

 {\subsection{Percolation on multiplex random hypergraphs}}
 
On a random multiplex hypergraph we can capture hypergraph percolation by 
introducing  
the probabilities $W_m$ that a node belonging to a random hyperedge in layer $m$ is  
occurs to be the root of an infinite tree when this hyperedge is deleted and the probability  $V_m$ that  a random hyperedge of a node in layer $m$ leads to the giant component.
These probabilities are determined by the self-consistency equations:
\bea
\!\!\!\!\!\!\!\!\!\!\!\!
W_m&=&\sum_{{\bf q}}\frac{q_m P({\bf q})}{\avg{q_m}}\left[1-\prod_{m'=2}^{M}(1-V_{m'})^{q_{m'}-\delta_{m,m'}}\right]
, 
\nonumber 
\\
\!\!\!\!\!\!\!\!\!\!\!\!
V_{m}&=&p_H^{[m]}p_N^{m-1}\left[1-(1-W_m)^{m-1}\right]
,
\eea
which are Eqs.~\eq{mp_1b} and \eq{mp_2b} averaged over the distribution $P(\vec{G})$ for a random multiplex hypergraph, discribed by the same Eq.~\eq{MultiplexG} as for the corresponding random multiplex factor graph.    
The probabilities $R$  and $S$ that, respectively, a node and an hyperedge are in the giant component of the multiplex hypergraph are given by 
\bea
R&=&p_N\sum_{{\bf q}}P({\bf q})\left[1-\prod_{m'=2}^{M}(1-V_{m'})^{q_{m'}}\right], 
\nonumber 
\\
S&=&\sum_{m}Q(m)p_H^{[m]}p_N^{m}\left[1-(1-W_m)^{m}\right]
, 
\eea
 {where for this model of a random hypergraph,  
\bea
Q(m) = \frac{\avg{m}}{m} \frac{\avg{q_m}}{\avg{q}}
.
\eea
}
These equations lead to a continuous second order phase transition for 
\bea
\Lambda(p_N,{\bf p}_H)=1
,
\eea
where $\Lambda(p_N,{\bf p}_H)$ is the maximum eigenvalue of the matrix ${\bf G}$ of elements
\bea
G_{m,m'}=p_H^{[m]}p_N^{m-1}(m-1)\frac{\avg{q_{m}(q_{m}-\delta_{m',m})}}{\avg{q_{m}}}
.
\label{5200}
\eea
The criterion of the existence of the giant connected component is this problem is $\Lambda(p_N,{\bf p}_H) > 1$. 
Inspecting the matrix elements in Eq.~\eq{5200}, we see that the node percolation threshold $p_{cN}$ in this %%uncorrelated 
random hypergraph coincides with the hyperedge percolation threshold if each hyperedge of cardinality $m$ is retained with probability $p_H^{[m]} = p_{cN}^{m-1} \leq p_{cN}$. 
%%This means that if 
If we remove hyperedges uniformly (undependently of their cardinalities), then the resulting hyperedge percolation threshold $p_{cH}$ must be lower than the maximum probability $p_H^{[m]} = p_{cN}^{m-1}$ for the nonuniform removal of edges, i.e. $p_{cH} < p_{cN}^{m_{\min}-1} \leq p_{cN}$. 
Thus the node percolation threshold for this model of an uncorrelated hypergraph exceeds the hyperedge percolation threshold, as we have already showed for general locally tree-like hypergraphs.

%%%%%%%%%%%%%%%%%%%%%%
%%%%%%%%%%%%%%%%%%%%%%
%%%%%%%%%%%%%%%%%%%%%%
%%%%%%%%%%%%%%%%%%%%%%

\section{Discussions and conclusion}
\label{s6}

We have formulated message-passing algorithms to investigate percolation on hypergraphs, we have studied the critical behavior of this process on arbitrary topologies and on two versions of the configuration model of random hypergraphs, and we have observed a large difference between the node and hyperedge percolation problems. The difference is particularly big when a hypergraph contains hyperedges with large cardinalities. We have shown that the node percolation threshold for  
a locally tree-like hypergraph exceeds the hyperedge percolation threshold. This qualitative difference between node and hyperedge percolation on hypergraphs is in marked contrast to ordinary networks, where these two types of percolation do not differ much from each other, and the node and hyperedge percolation thresholds coincide for a locally tree-like network.  
 {Moreover, if the second moment of the degree distribution of nodes in a hypergraph is finite, then its node percolation threshold is finite for any cardinality distribution of hyperedges, even with the first moment diverging. That is, any fat-tailed cardinality distribution cannot lead to the hyper-resilience phenomenon in hypergraphs in contrast to their factor graphs, where the divergent second moment of a cardinality distribution guarantees zero percolation threshold and the hyper-resilience phenomenon.}

The node percolation problem is the basic problem in which results for bipartite networks cannot be directly applied to their hypergraph counterparts. We have shown that the node percolation threshold of a locally tree-like hypergraph exceeds this threshold for the corresponding factor graph. We have observed this effect also in a sparse real-world hypergraph whose structure is not tree-like. One can indicate a number of other problems and processes, where such a direct mapping is impossible. These problems involve the removal of nodes. This occurs, in particular, during various pruning processes, including the $k$-core pruning process, cascading failures in multilayer interdependent networks, etc., and during disease spreading in various epidemic models. 

Importantly, our equations of the message-passing algorithm do not assume the absence of correlations in hypergraphs. They allow a numerical treatment of the problem. The final formulas have been obtained for 
hypergraphs having no degree-degree correlations between different nodes. We suggest that an analytical treatment of correlated hypergraphs is also possible in the framework of our approach.

%%%%%%%%%%%%%%%%%%%%%%%%%%%%%%%%%%%%%%%%%%%%%%%%%%%%%%%%%%%%%%%%%%%%%%%
%%%%%%%%%%%%%%%%%%%%%%%%%%%%%%%%%%%%%%%%%%%%%%%%%%%%%%%%%%%%%%%%%%%%%%%

%%%%%%%%%%%%%%%%%%%%%%%%%%%%%%%%%%%%%%%%%%%%%%%%%%%%%%%%%%%%%%%%%%%%%%%
%%%%%%%%%%%%%%%%%%%%%%%%%%%%%%%%%%%%%%%%%%%%%%%%%%%%%%%%%%%%%%%%%%%%%%%

%%%%%%%%%%%%%%%%%%%%%%%%%%%%%%%%%%%%%%%%%%%%%%%%%%%%%%%%%%%%%%%%%%%%%%%
%%%%%%%%%%%%%%%%%%%%%%%%%%%%%%%%%%%%%%%%%%%%%%%%%%%%%%%%%%%%%%%%%%%%%%%

%%%%%%%%%%%
%%%%%%%%%%%
%%%%%%%%%%%
%%%%%%%%%%%

\bibliography{site_percolation_bibliography}

\begin{thebibliography}{50}%
\makeatletter
\providecommand \@ifxundefined [1]{%
 \@ifx{#1\undefined}
}%
\providecommand \@ifnum [1]{%
 \ifnum #1\expandafter \@firstoftwo
 \else \expandafter \@secondoftwo
 \fi
}%
\providecommand \@ifx [1]{%
 \ifx #1\expandafter \@firstoftwo
 \else \expandafter \@secondoftwo
 \fi
}%
\providecommand \natexlab [1]{#1}%
\providecommand \enquote  [1]{``#1''}%
\providecommand \bibnamefont  [1]{#1}%
\providecommand \bibfnamefont [1]{#1}%
\providecommand \citenamefont [1]{#1}%
\providecommand \href@noop [0]{\@secondoftwo}%
\providecommand \href [0]{\begingroup \@sanitize@url \@href}%
\providecommand \@href[1]{\@@startlink{#1}\@@href}%
\providecommand \@@href[1]{\endgroup#1\@@endlink}%
\providecommand \@sanitize@url [0]{\catcode `\\12\catcode `\$12\catcode
  `\&12\catcode `\#12\catcode `\^12\catcode `\_12\catcode `\%12\relax}%
\providecommand \@@startlink[1]{}%
\providecommand \@@endlink[0]{}%
\providecommand \url  [0]{\begingroup\@sanitize@url \@url }%
\providecommand \@url [1]{\endgroup\@href {#1}{\urlprefix }}%
\providecommand \urlprefix  [0]{URL }%
\providecommand \Eprint [0]{\href }%
\providecommand \doibase [0]{https://doi.org/}%
\providecommand \selectlanguage [0]{\@gobble}%
\providecommand \bibinfo  [0]{\@secondoftwo}%
\providecommand \bibfield  [0]{\@secondoftwo}%
\providecommand \translation [1]{[#1]}%
\providecommand \BibitemOpen [0]{}%
\providecommand \bibitemStop [0]{}%
\providecommand \bibitemNoStop [0]{.\EOS\space}%
\providecommand \EOS [0]{\spacefactor3000\relax}%
\providecommand \BibitemShut  [1]{\csname bibitem#1\endcsname}%
\let\auto@bib@innerbib\@empty
%</preamble>
\bibitem [{\citenamefont {Benson}\ \emph {et~al.}(2016)\citenamefont {Benson},
  \citenamefont {Gleich},\ and\ \citenamefont {Leskovec}}]{benson2016higher}%
  \BibitemOpen
  \bibfield  {author} {\bibinfo {author} {\bibfnamefont {A.~R.}\ \bibnamefont
  {Benson}}, \bibinfo {author} {\bibfnamefont {D.~F.}\ \bibnamefont {Gleich}},\
  and\ \bibinfo {author} {\bibfnamefont {J.}~\bibnamefont {Leskovec}},\
  }\bibfield  {title} {\bibinfo {title} {Higher-order organization of complex
  networks},\ }\href@noop {} {\bibfield  {journal} {\bibinfo  {journal}
  {Science}\ }\textbf {\bibinfo {volume} {353}},\ \bibinfo {pages} {163}
  (\bibinfo {year} {2016})}\BibitemShut {NoStop}%
\bibitem [{\citenamefont {Battiston}\ \emph {et~al.}(2020)\citenamefont
  {Battiston}, \citenamefont {Cencetti}, \citenamefont {Iacopini},
  \citenamefont {Latora}, \citenamefont {Lucas}, \citenamefont {Patania},
  \citenamefont {Young},\ and\ \citenamefont {Petri}}]{battiston2020networks}%
  \BibitemOpen
  \bibfield  {author} {\bibinfo {author} {\bibfnamefont {F.}~\bibnamefont
  {Battiston}}, \bibinfo {author} {\bibfnamefont {G.}~\bibnamefont {Cencetti}},
  \bibinfo {author} {\bibfnamefont {I.}~\bibnamefont {Iacopini}}, \bibinfo
  {author} {\bibfnamefont {V.}~\bibnamefont {Latora}}, \bibinfo {author}
  {\bibfnamefont {M.}~\bibnamefont {Lucas}}, \bibinfo {author} {\bibfnamefont
  {A.}~\bibnamefont {Patania}}, \bibinfo {author} {\bibfnamefont {J.-G.}\
  \bibnamefont {Young}},\ and\ \bibinfo {author} {\bibfnamefont
  {G.}~\bibnamefont {Petri}},\ }\bibfield  {title} {\bibinfo {title} {{Networks
  beyond pairwise interactions: Structure and dynamics}},\ }\href@noop {}
  {\bibfield  {journal} {\bibinfo  {journal} {Phys. Rep.}\ }\textbf {\bibinfo
  {volume} {874}},\ \bibinfo {pages} {1} (\bibinfo {year} {2020})}\BibitemShut
  {NoStop}%
\bibitem [{\citenamefont {Bianconi}(2021)}]{bianconi2021higher}%
  \BibitemOpen
  \bibfield  {author} {\bibinfo {author} {\bibfnamefont {G.}~\bibnamefont
  {Bianconi}},\ }\href@noop {} {\emph {\bibinfo {title} {Higher-Order
  Networks}}}\ (\bibinfo  {publisher} {Cambridge University Press, Cambridge},\
  \bibinfo {year} {2021})\BibitemShut {NoStop}%
\bibitem [{\citenamefont {Boccaletti}\ \emph {et~al.}(2023)\citenamefont
  {Boccaletti}, \citenamefont {De~Lellis}, \citenamefont {del Genio},
  \citenamefont {Alfaro-Bittner}, \citenamefont {Criado}, \citenamefont
  {Jalan},\ and\ \citenamefont {Romance}}]{boccaletti2023structure}%
  \BibitemOpen
  \bibfield  {author} {\bibinfo {author} {\bibfnamefont {S.}~\bibnamefont
  {Boccaletti}}, \bibinfo {author} {\bibfnamefont {P.}~\bibnamefont
  {De~Lellis}}, \bibinfo {author} {\bibfnamefont {C.~I.}\ \bibnamefont {del
  Genio}}, \bibinfo {author} {\bibfnamefont {K.}~\bibnamefont
  {Alfaro-Bittner}}, \bibinfo {author} {\bibfnamefont {R.}~\bibnamefont
  {Criado}}, \bibinfo {author} {\bibfnamefont {S.}~\bibnamefont {Jalan}},\ and\
  \bibinfo {author} {\bibfnamefont {M.}~\bibnamefont {Romance}},\ }\bibfield
  {title} {\bibinfo {title} {The structure and dynamics of networks with higher
  order interactions},\ }\href@noop {} {\bibfield  {journal} {\bibinfo
  {journal} {Phys. Rep.}\ }\textbf {\bibinfo {volume} {1018}},\ \bibinfo
  {pages} {1} (\bibinfo {year} {2023})}\BibitemShut {NoStop}%
\bibitem [{\citenamefont {Torres}\ \emph {et~al.}(2021)\citenamefont {Torres},
  \citenamefont {Blevins}, \citenamefont {Bassett},\ and\ \citenamefont
  {Eliassi-Rad}}]{torres2021and}%
  \BibitemOpen
  \bibfield  {author} {\bibinfo {author} {\bibfnamefont {L.}~\bibnamefont
  {Torres}}, \bibinfo {author} {\bibfnamefont {A.~S.}\ \bibnamefont {Blevins}},
  \bibinfo {author} {\bibfnamefont {D.}~\bibnamefont {Bassett}},\ and\ \bibinfo
  {author} {\bibfnamefont {T.}~\bibnamefont {Eliassi-Rad}},\ }\bibfield
  {title} {\bibinfo {title} {The why, how, and when of representations for
  complex systems},\ }\href@noop {} {\bibfield  {journal} {\bibinfo  {journal}
  {SIAM Rev.}\ }\textbf {\bibinfo {volume} {63}},\ \bibinfo {pages} {435}
  (\bibinfo {year} {2021})}\BibitemShut {NoStop}%
\bibitem [{\citenamefont {Bianconi}(2022)}]{bianconi2022statistical}%
  \BibitemOpen
  \bibfield  {author} {\bibinfo {author} {\bibfnamefont {G.}~\bibnamefont
  {Bianconi}},\ }\bibfield  {title} {\bibinfo {title} {Statistical physics of
  exchangeable sparse simple networks, multiplex networks, and simplicial
  complexes},\ }\href@noop {} {\bibfield  {journal} {\bibinfo  {journal} {Phys.
  Rev. E}\ }\textbf {\bibinfo {volume} {105}},\ \bibinfo {pages} {034310}
  (\bibinfo {year} {2022})}\BibitemShut {NoStop}%
\bibitem [{\citenamefont {Barthelemy}(2022)}]{barthelemy2022class}%
  \BibitemOpen
  \bibfield  {author} {\bibinfo {author} {\bibfnamefont {M.}~\bibnamefont
  {Barthelemy}},\ }\bibfield  {title} {\bibinfo {title} {Class of models for
  random hypergraphs},\ }\href@noop {} {\bibfield  {journal} {\bibinfo
  {journal} {Phys. Rev. E}\ }\textbf {\bibinfo {volume} {106}},\ \bibinfo
  {pages} {064310} (\bibinfo {year} {2022})}\BibitemShut {NoStop}%
\bibitem [{\citenamefont {Krapivsky}(2023)}]{krapivsky2023random}%
  \BibitemOpen
  \bibfield  {author} {\bibinfo {author} {\bibfnamefont {P.~L.}\ \bibnamefont
  {Krapivsky}},\ }\bibfield  {title} {\bibinfo {title} {Random recursive
  hypergraphs},\ }\href@noop {} {\bibfield  {journal} {\bibinfo  {journal} {J.
  Phys. A: Mathematical and Theoretical}\ }\textbf {\bibinfo {volume} {56}},\
  \bibinfo {pages} {195001} (\bibinfo {year} {2023})}\BibitemShut {NoStop}%
\bibitem [{\citenamefont {Battiston}\ \emph {et~al.}(2021)\citenamefont
  {Battiston}, \citenamefont {Amico}, \citenamefont {Barrat}, \citenamefont
  {Bianconi}, \citenamefont {Ferraz~de Arruda}, \citenamefont {Franceschiello},
  \citenamefont {Iacopini}, \citenamefont {K{\'e}fi}, \citenamefont {Latora},
  \citenamefont {Moreno}, \citenamefont {Murray}, \citenamefont {Peixoto},
  \citenamefont {Vaccarino},\ and\ \citenamefont
  {Petri}}]{battiston2021physics}%
  \BibitemOpen
  \bibfield  {author} {\bibinfo {author} {\bibfnamefont {F.}~\bibnamefont
  {Battiston}}, \bibinfo {author} {\bibfnamefont {E.}~\bibnamefont {Amico}},
  \bibinfo {author} {\bibfnamefont {A.}~\bibnamefont {Barrat}}, \bibinfo
  {author} {\bibfnamefont {G.}~\bibnamefont {Bianconi}}, \bibinfo {author}
  {\bibfnamefont {G.}~\bibnamefont {Ferraz~de Arruda}}, \bibinfo {author}
  {\bibfnamefont {B.}~\bibnamefont {Franceschiello}}, \bibinfo {author}
  {\bibfnamefont {I.}~\bibnamefont {Iacopini}}, \bibinfo {author}
  {\bibfnamefont {S.}~\bibnamefont {K{\'e}fi}}, \bibinfo {author}
  {\bibfnamefont {V.}~\bibnamefont {Latora}}, \bibinfo {author} {\bibfnamefont
  {Y.}~\bibnamefont {Moreno}}, \bibinfo {author} {\bibfnamefont {M.~M.}\
  \bibnamefont {Murray}}, \bibinfo {author} {\bibfnamefont {T.~P.}\
  \bibnamefont {Peixoto}}, \bibinfo {author} {\bibfnamefont {F.}~\bibnamefont
  {Vaccarino}},\ and\ \bibinfo {author} {\bibfnamefont {G.}~\bibnamefont
  {Petri}},\ }\bibfield  {title} {\bibinfo {title} {The physics of higher-order
  interactions in complex systems},\ }\href@noop {} {\bibfield  {journal}
  {\bibinfo  {journal} {Nature Phys.}\ }\textbf {\bibinfo {volume} {17}},\
  \bibinfo {pages} {1093} (\bibinfo {year} {2021})}\BibitemShut {NoStop}%
\bibitem [{\citenamefont {Majhi}\ \emph {et~al.}(2022)\citenamefont {Majhi},
  \citenamefont {Perc},\ and\ \citenamefont {Ghosh}}]{majhi2022dynamics}%
  \BibitemOpen
  \bibfield  {author} {\bibinfo {author} {\bibfnamefont {S.}~\bibnamefont
  {Majhi}}, \bibinfo {author} {\bibfnamefont {M.}~\bibnamefont {Perc}},\ and\
  \bibinfo {author} {\bibfnamefont {D.}~\bibnamefont {Ghosh}},\ }\bibfield
  {title} {\bibinfo {title} {{Dynamics on higher-order networks: A review}},\
  }\href@noop {} {\bibfield  {journal} {\bibinfo  {journal} {J. Royal Soc.
  Interface}\ }\textbf {\bibinfo {volume} {19}},\ \bibinfo {pages} {20220043}
  (\bibinfo {year} {2022})}\BibitemShut {NoStop}%
\bibitem [{\citenamefont {Coutinho}\ \emph {et~al.}(2020)\citenamefont
  {Coutinho}, \citenamefont {Wu}, \citenamefont {Zhou},\ and\ \citenamefont
  {Liu}}]{coutinho2020covering}%
  \BibitemOpen
  \bibfield  {author} {\bibinfo {author} {\bibfnamefont {B.~C.}\ \bibnamefont
  {Coutinho}}, \bibinfo {author} {\bibfnamefont {A.-K.}\ \bibnamefont {Wu}},
  \bibinfo {author} {\bibfnamefont {H.-J.}\ \bibnamefont {Zhou}},\ and\
  \bibinfo {author} {\bibfnamefont {Y.-Y.}\ \bibnamefont {Liu}},\ }\bibfield
  {title} {\bibinfo {title} {Covering problems and core percolations on
  hypergraphs},\ }\href@noop {} {\bibfield  {journal} {\bibinfo  {journal}
  {Phys. Rev. Letts.}\ }\textbf {\bibinfo {volume} {124}},\ \bibinfo {pages}
  {248301} (\bibinfo {year} {2020})}\BibitemShut {NoStop}%
\bibitem [{\citenamefont {Sun}\ and\ \citenamefont
  {Bianconi}(2021)}]{sun2021higher}%
  \BibitemOpen
  \bibfield  {author} {\bibinfo {author} {\bibfnamefont {H.}~\bibnamefont
  {Sun}}\ and\ \bibinfo {author} {\bibfnamefont {G.}~\bibnamefont {Bianconi}},\
  }\bibfield  {title} {\bibinfo {title} {Higher-order percolation processes on
  multiplex hypergraphs},\ }\href@noop {} {\bibfield  {journal} {\bibinfo
  {journal} {Phys. Rev. E}\ }\textbf {\bibinfo {volume} {104}},\ \bibinfo
  {pages} {034306} (\bibinfo {year} {2021})}\BibitemShut {NoStop}%
\bibitem [{\citenamefont {Sun}\ \emph {et~al.}(2023)\citenamefont {Sun},
  \citenamefont {Radicchi}, \citenamefont {Kurths},\ and\ \citenamefont
  {Bianconi}}]{sun2023dynamic}%
  \BibitemOpen
  \bibfield  {author} {\bibinfo {author} {\bibfnamefont {H.}~\bibnamefont
  {Sun}}, \bibinfo {author} {\bibfnamefont {F.}~\bibnamefont {Radicchi}},
  \bibinfo {author} {\bibfnamefont {J.}~\bibnamefont {Kurths}},\ and\ \bibinfo
  {author} {\bibfnamefont {G.}~\bibnamefont {Bianconi}},\ }\bibfield  {title}
  {\bibinfo {title} {The dynamic nature of percolation on networks with triadic
  interactions},\ }\href@noop {} {\bibfield  {journal} {\bibinfo  {journal}
  {Nature Commun.}\ }\textbf {\bibinfo {volume} {14}},\ \bibinfo {pages} {1308}
  (\bibinfo {year} {2023})}\BibitemShut {NoStop}%
\bibitem [{\citenamefont {Lee}\ \emph {et~al.}(2023)\citenamefont {Lee},
  \citenamefont {Goh}, \citenamefont {Lee},\ and\ \citenamefont
  {Kahng}}]{lee2023k}%
  \BibitemOpen
  \bibfield  {author} {\bibinfo {author} {\bibfnamefont {J.}~\bibnamefont
  {Lee}}, \bibinfo {author} {\bibfnamefont {K.-I.}\ \bibnamefont {Goh}},
  \bibinfo {author} {\bibfnamefont {D.-S.}\ \bibnamefont {Lee}},\ and\ \bibinfo
  {author} {\bibfnamefont {B.}~\bibnamefont {Kahng}},\ }\bibfield  {title}
  {\bibinfo {title} {$(k, q)$-core decomposition of hypergraphs},\ }\href@noop
  {} {\bibfield  {journal} {\bibinfo  {journal} {arXiv:2301.06712}\ } (\bibinfo
  {year} {2023})}\BibitemShut {NoStop}%
\bibitem [{\citenamefont {Peng}\ \emph {et~al.}(2022)\citenamefont {Peng},
  \citenamefont {Qian}, \citenamefont {Zhao}, \citenamefont {Zhong},
  \citenamefont {Ling},\ and\ \citenamefont {Wang}}]{peng2022disintegrate}%
  \BibitemOpen
  \bibfield  {author} {\bibinfo {author} {\bibfnamefont {H.}~\bibnamefont
  {Peng}}, \bibinfo {author} {\bibfnamefont {C.}~\bibnamefont {Qian}}, \bibinfo
  {author} {\bibfnamefont {D.}~\bibnamefont {Zhao}}, \bibinfo {author}
  {\bibfnamefont {M.}~\bibnamefont {Zhong}}, \bibinfo {author} {\bibfnamefont
  {X.}~\bibnamefont {Ling}},\ and\ \bibinfo {author} {\bibfnamefont
  {W.}~\bibnamefont {Wang}},\ }\bibfield  {title} {\bibinfo {title}
  {Disintegrate hypergraph networks by attacking hyperedge},\ }\href@noop {}
  {\bibfield  {journal} {\bibinfo  {journal} {Journal of King Saud
  University-Computer and Information Sciences}\ }\textbf {\bibinfo {volume}
  {34}},\ \bibinfo {pages} {4679} (\bibinfo {year} {2022})}\BibitemShut
  {NoStop}%
\bibitem [{\citenamefont {Peng}\ \emph {et~al.}(2023)\citenamefont {Peng},
  \citenamefont {Qian}, \citenamefont {Zhao}, \citenamefont {Zhong},
  \citenamefont {Han}, \citenamefont {Li},\ and\ \citenamefont
  {Wang}}]{peng2023message}%
  \BibitemOpen
  \bibfield  {author} {\bibinfo {author} {\bibfnamefont {H.}~\bibnamefont
  {Peng}}, \bibinfo {author} {\bibfnamefont {C.}~\bibnamefont {Qian}}, \bibinfo
  {author} {\bibfnamefont {D.}~\bibnamefont {Zhao}}, \bibinfo {author}
  {\bibfnamefont {M.}~\bibnamefont {Zhong}}, \bibinfo {author} {\bibfnamefont
  {J.}~\bibnamefont {Han}}, \bibinfo {author} {\bibfnamefont {R.}~\bibnamefont
  {Li}},\ and\ \bibinfo {author} {\bibfnamefont {W.}~\bibnamefont {Wang}},\
  }\bibfield  {title} {\bibinfo {title} {Message passing approach to analyze
  the robustness of hypergraph},\ }\href@noop {} {\bibfield  {journal}
  {\bibinfo  {journal} {arXiv:2302.14594}\ } (\bibinfo {year}
  {2023})}\BibitemShut {NoStop}%
\bibitem [{\citenamefont {Mancastroppa}\ \emph {et~al.}(2023)\citenamefont
  {Mancastroppa}, \citenamefont {Iacopini}, \citenamefont {Petri},\ and\
  \citenamefont {Barrat}}]{mancastroppa2023hyper}%
  \BibitemOpen
  \bibfield  {author} {\bibinfo {author} {\bibfnamefont {M.}~\bibnamefont
  {Mancastroppa}}, \bibinfo {author} {\bibfnamefont {I.}~\bibnamefont
  {Iacopini}}, \bibinfo {author} {\bibfnamefont {G.}~\bibnamefont {Petri}},\
  and\ \bibinfo {author} {\bibfnamefont {A.}~\bibnamefont {Barrat}},\
  }\bibfield  {title} {\bibinfo {title} {Hyper-cores promote localization and
  efficient seeding in higher-order processes},\ }\href@noop {} {\bibfield
  {journal} {\bibinfo  {journal} {Nature Communications}\ }\textbf {\bibinfo
  {volume} {14}},\ \bibinfo {pages} {62223} (\bibinfo {year}
  {2023})}\BibitemShut {NoStop}%
\bibitem [{\citenamefont {Thurner}\ \emph {et~al.}(2010)\citenamefont
  {Thurner}, \citenamefont {Klimek},\ and\ \citenamefont
  {Hanel}}]{thurner2010schumpeterian}%
  \BibitemOpen
  \bibfield  {author} {\bibinfo {author} {\bibfnamefont {S.}~\bibnamefont
  {Thurner}}, \bibinfo {author} {\bibfnamefont {P.}~\bibnamefont {Klimek}},\
  and\ \bibinfo {author} {\bibfnamefont {R.}~\bibnamefont {Hanel}},\ }\bibfield
   {title} {\bibinfo {title} {Schumpeterian economic dynamics as a quantifiable
  model of evolution},\ }\href@noop {} {\bibfield  {journal} {\bibinfo
  {journal} {New J. Phys.}\ }\textbf {\bibinfo {volume} {12}},\ \bibinfo
  {pages} {075029} (\bibinfo {year} {2010})}\BibitemShut {NoStop}%
\bibitem [{\citenamefont {Hanel}\ \emph {et~al.}(2005)\citenamefont {Hanel},
  \citenamefont {Kauffman},\ and\ \citenamefont {Thurner}}]{hanel2005phase}%
  \BibitemOpen
  \bibfield  {author} {\bibinfo {author} {\bibfnamefont {R.}~\bibnamefont
  {Hanel}}, \bibinfo {author} {\bibfnamefont {S.~A.}\ \bibnamefont
  {Kauffman}},\ and\ \bibinfo {author} {\bibfnamefont {S.}~\bibnamefont
  {Thurner}},\ }\bibfield  {title} {\bibinfo {title} {Phase transition in
  random catalytic networks},\ }\href@noop {} {\bibfield  {journal} {\bibinfo
  {journal} {Phys. Rev. E}\ }\textbf {\bibinfo {volume} {72}},\ \bibinfo
  {pages} {036117} (\bibinfo {year} {2005})}\BibitemShut {NoStop}%
\bibitem [{\citenamefont {Klimm}\ \emph {et~al.}(2021)\citenamefont {Klimm},
  \citenamefont {Deane},\ and\ \citenamefont {Reinert}}]{klimm2021hypergraphs}%
  \BibitemOpen
  \bibfield  {author} {\bibinfo {author} {\bibfnamefont {F.}~\bibnamefont
  {Klimm}}, \bibinfo {author} {\bibfnamefont {C.~M.}\ \bibnamefont {Deane}},\
  and\ \bibinfo {author} {\bibfnamefont {G.}~\bibnamefont {Reinert}},\
  }\bibfield  {title} {\bibinfo {title} {Hypergraphs for predicting essential
  genes using multiprotein complex data},\ }\href@noop {} {\bibfield  {journal}
  {\bibinfo  {journal} {J. Complex Networks}\ }\textbf {\bibinfo {volume}
  {9}},\ \bibinfo {pages} {cnaa028} (\bibinfo {year} {2021})}\BibitemShut
  {NoStop}%
\bibitem [{\citenamefont {Jost}\ and\ \citenamefont
  {Mulas}(2019)}]{jost2019hypergraph}%
  \BibitemOpen
  \bibfield  {author} {\bibinfo {author} {\bibfnamefont {J.}~\bibnamefont
  {Jost}}\ and\ \bibinfo {author} {\bibfnamefont {R.}~\bibnamefont {Mulas}},\
  }\bibfield  {title} {\bibinfo {title} {{Hypergraph Laplace operators for
  chemical reaction networks}},\ }\href@noop {} {\bibfield  {journal} {\bibinfo
   {journal} {Adv. Math.}\ }\textbf {\bibinfo {volume} {351}},\ \bibinfo
  {pages} {870} (\bibinfo {year} {2019})}\BibitemShut {NoStop}%
\bibitem [{\citenamefont
  {Bianconi}(2018{\natexlab{a}})}]{bianconi2018multilayer}%
  \BibitemOpen
  \bibfield  {author} {\bibinfo {author} {\bibfnamefont {G.}~\bibnamefont
  {Bianconi}},\ }\href@noop {} {\emph {\bibinfo {title} {Multilayer Networks:
  Structure and Function}}}\ (\bibinfo  {publisher} {Oxford University Press},\
  \bibinfo {address} {Oxford},\ \bibinfo {year} {2018})\BibitemShut {NoStop}%
\bibitem [{\citenamefont {Newman}(2023)}]{newman2023message}%
  \BibitemOpen
  \bibfield  {author} {\bibinfo {author} {\bibfnamefont {M.~E.~J.}\
  \bibnamefont {Newman}},\ }\bibfield  {title} {\bibinfo {title} {Message
  passing methods on complex networks},\ }\href@noop {} {\bibfield  {journal}
  {\bibinfo  {journal} {Proc. Royal Soc. A}\ }\textbf {\bibinfo {volume}
  {479}},\ \bibinfo {pages} {20220774} (\bibinfo {year} {2023})}\BibitemShut
  {NoStop}%
\bibitem [{\citenamefont {Hartmann}\ and\ \citenamefont
  {Weigt}(2006)}]{hartmann2006phase}%
  \BibitemOpen
  \bibfield  {author} {\bibinfo {author} {\bibfnamefont {A.~K.}\ \bibnamefont
  {Hartmann}}\ and\ \bibinfo {author} {\bibfnamefont {M.}~\bibnamefont
  {Weigt}},\ }\href@noop {} {\emph {\bibinfo {title} {Phase Transitions in
  Combinatorial Optimization Problems: Basics, Algorithms and Statistical
  Mechanics}}}\ (\bibinfo  {publisher} {John Wiley \& Sons, Weinheim},\
  \bibinfo {year} {2006})\BibitemShut {NoStop}%
\bibitem [{\citenamefont {Karrer}\ and\ \citenamefont
  {Newman}(2010)}]{karrer2010message}%
  \BibitemOpen
  \bibfield  {author} {\bibinfo {author} {\bibfnamefont {B.}~\bibnamefont
  {Karrer}}\ and\ \bibinfo {author} {\bibfnamefont {M.~E.~J.}\ \bibnamefont
  {Newman}},\ }\bibfield  {title} {\bibinfo {title} {Message passing approach
  for general epidemic models},\ }\href@noop {} {\bibfield  {journal} {\bibinfo
   {journal} {Phys. Rev. E}\ }\textbf {\bibinfo {volume} {82}},\ \bibinfo
  {pages} {016101} (\bibinfo {year} {2010})}\BibitemShut {NoStop}%
\bibitem [{\citenamefont {Zhao}\ and\ \citenamefont
  {Bianconi}(2013)}]{zhao2013antagonistic}%
  \BibitemOpen
  \bibfield  {author} {\bibinfo {author} {\bibfnamefont {K.}~\bibnamefont
  {Zhao}}\ and\ \bibinfo {author} {\bibfnamefont {G.}~\bibnamefont
  {Bianconi}},\ }\bibfield  {title} {\bibinfo {title} {Percolation on
  interacting, antagonistic networks},\ }\href@noop {} {\bibfield  {journal}
  {\bibinfo  {journal} {J. Stat. Mech.: Theory and Experiment}\ }\textbf
  {\bibinfo {volume} {2013}},\ \bibinfo {pages} {P05005} (\bibinfo {year}
  {2013})}\BibitemShut {NoStop}%
\bibitem [{\citenamefont {Cellai}\ \emph {et~al.}(2016)\citenamefont {Cellai},
  \citenamefont {Dorogovtsev},\ and\ \citenamefont
  {Bianconi}}]{cellai2016message}%
  \BibitemOpen
  \bibfield  {author} {\bibinfo {author} {\bibfnamefont {D.}~\bibnamefont
  {Cellai}}, \bibinfo {author} {\bibfnamefont {S.~N.}\ \bibnamefont
  {Dorogovtsev}},\ and\ \bibinfo {author} {\bibfnamefont {G.}~\bibnamefont
  {Bianconi}},\ }\bibfield  {title} {\bibinfo {title} {Message passing theory
  for percolation models on multiplex networks with link overlap},\ }\href@noop
  {} {\bibfield  {journal} {\bibinfo  {journal} {Phys. Rev. E}\ }\textbf
  {\bibinfo {volume} {94}},\ \bibinfo {pages} {032301} (\bibinfo {year}
  {2016})}\BibitemShut {NoStop}%
\bibitem [{\citenamefont {Radicchi}\ and\ \citenamefont
  {Bianconi}(2017)}]{radicchi2017redundant}%
  \BibitemOpen
  \bibfield  {author} {\bibinfo {author} {\bibfnamefont {F.}~\bibnamefont
  {Radicchi}}\ and\ \bibinfo {author} {\bibfnamefont {G.}~\bibnamefont
  {Bianconi}},\ }\bibfield  {title} {\bibinfo {title} {Redundant
  interdependencies boost the robustness of multiplex networks},\ }\href@noop
  {} {\bibfield  {journal} {\bibinfo  {journal} {Phys. Rev. X}\ }\textbf
  {\bibinfo {volume} {7}},\ \bibinfo {pages} {011013} (\bibinfo {year}
  {2017})}\BibitemShut {NoStop}%
\bibitem [{\citenamefont {Watanabe}\ and\ \citenamefont
  {Kabashima}(2014)}]{watanabe2014cavity}%
  \BibitemOpen
  \bibfield  {author} {\bibinfo {author} {\bibfnamefont {S.}~\bibnamefont
  {Watanabe}}\ and\ \bibinfo {author} {\bibfnamefont {Y.}~\bibnamefont
  {Kabashima}},\ }\bibfield  {title} {\bibinfo {title} {{Cavity-based
  robustness analysis of interdependent networks: Influences of intranetwork
  and internetwork degree-degree correlations}},\ }\href@noop {} {\bibfield
  {journal} {\bibinfo  {journal} {Phys. Rev. E}\ }\textbf {\bibinfo {volume}
  {89}},\ \bibinfo {pages} {012808} (\bibinfo {year} {2014})}\BibitemShut
  {NoStop}%
\bibitem [{\citenamefont {Cantwell}\ \emph {et~al.}(2023)\citenamefont
  {Cantwell}, \citenamefont {Kirkley},\ and\ \citenamefont
  {Radicchi}}]{cantwell2023heterogeneous}%
  \BibitemOpen
  \bibfield  {author} {\bibinfo {author} {\bibfnamefont {G.~T.}\ \bibnamefont
  {Cantwell}}, \bibinfo {author} {\bibfnamefont {A.}~\bibnamefont {Kirkley}},\
  and\ \bibinfo {author} {\bibfnamefont {F.}~\bibnamefont {Radicchi}},\
  }\bibfield  {title} {\bibinfo {title} {Heterogeneous message passing for
  heterogeneous networks},\ }\href@noop {} {\bibfield  {journal} {\bibinfo
  {journal} {arXiv:2305.02294}\ } (\bibinfo {year} {2023})}\BibitemShut
  {NoStop}%
\bibitem [{\citenamefont {Bianconi}(2018{\natexlab{b}})}]{bianconi2018rare}%
  \BibitemOpen
  \bibfield  {author} {\bibinfo {author} {\bibfnamefont {G.}~\bibnamefont
  {Bianconi}},\ }\bibfield  {title} {\bibinfo {title} {Rare events and
  discontinuous percolation transitions},\ }\href@noop {} {\bibfield  {journal}
  {\bibinfo  {journal} {Phys. Rev. E}\ }\textbf {\bibinfo {volume} {97}},\
  \bibinfo {pages} {022314} (\bibinfo {year} {2018}{\natexlab{b}})}\BibitemShut
  {NoStop}%
\bibitem [{\citenamefont {Cantwell}\ and\ \citenamefont
  {Newman}(2019)}]{cantwell2019message}%
  \BibitemOpen
  \bibfield  {author} {\bibinfo {author} {\bibfnamefont {G.~T.}\ \bibnamefont
  {Cantwell}}\ and\ \bibinfo {author} {\bibfnamefont {M.~E.~J.}\ \bibnamefont
  {Newman}},\ }\bibfield  {title} {\bibinfo {title} {Message passing on
  networks with loops},\ }\href@noop {} {\bibfield  {journal} {\bibinfo
  {journal} {PNAS}\ }\textbf {\bibinfo {volume} {116}},\ \bibinfo {pages}
  {23398} (\bibinfo {year} {2019})}\BibitemShut {NoStop}%
\bibitem [{\citenamefont {Mezard}\ and\ \citenamefont
  {Montanari}(2009)}]{mezard2009information}%
  \BibitemOpen
  \bibfield  {author} {\bibinfo {author} {\bibfnamefont {M.}~\bibnamefont
  {Mezard}}\ and\ \bibinfo {author} {\bibfnamefont {A.}~\bibnamefont
  {Montanari}},\ }\href@noop {} {\emph {\bibinfo {title} {Information, Physics,
  and Computation}}}\ (\bibinfo  {publisher} {Oxford University Press,
  Oxford},\ \bibinfo {year} {2009})\BibitemShut {NoStop}%
\bibitem [{\citenamefont {Yoon}\ \emph {et~al.}(2011)\citenamefont {Yoon},
  \citenamefont {Goltsev}, \citenamefont {Dorogovtsev},\ and\ \citenamefont
  {Mendes}}]{yoon2011belief}%
  \BibitemOpen
  \bibfield  {author} {\bibinfo {author} {\bibfnamefont {S.}~\bibnamefont
  {Yoon}}, \bibinfo {author} {\bibfnamefont {A.~V.}\ \bibnamefont {Goltsev}},
  \bibinfo {author} {\bibfnamefont {S.~N.}\ \bibnamefont {Dorogovtsev}},\ and\
  \bibinfo {author} {\bibfnamefont {J.~F.~F.}\ \bibnamefont {Mendes}},\
  }\bibfield  {title} {\bibinfo {title} {{Belief-propagation algorithm and the
  Ising model on networks with arbitrary distributions of motifs}},\
  }\href@noop {} {\bibfield  {journal} {\bibinfo  {journal} {Phys. Rev. E}\
  }\textbf {\bibinfo {volume} {84}},\ \bibinfo {pages} {041144} (\bibinfo
  {year} {2011})}\BibitemShut {NoStop}%
\bibitem [{\citenamefont {Sun}\ \emph {et~al.}(2021)\citenamefont {Sun},
  \citenamefont {Saad},\ and\ \citenamefont {Lokhov}}]{sun2021competition}%
  \BibitemOpen
  \bibfield  {author} {\bibinfo {author} {\bibfnamefont {H.}~\bibnamefont
  {Sun}}, \bibinfo {author} {\bibfnamefont {D.}~\bibnamefont {Saad}},\ and\
  \bibinfo {author} {\bibfnamefont {A.~Y.}\ \bibnamefont {Lokhov}},\ }\bibfield
   {title} {\bibinfo {title} {Competition, collaboration, and optimization in
  multiple interacting spreading processes},\ }\href@noop {} {\bibfield
  {journal} {\bibinfo  {journal} {Phys. Rev. X}\ }\textbf {\bibinfo {volume}
  {11}},\ \bibinfo {pages} {011048} (\bibinfo {year} {2021})}\BibitemShut
  {NoStop}%
\bibitem [{\citenamefont {Liu}\ \emph {et~al.}(2011)\citenamefont {Liu},
  \citenamefont {Slotine},\ and\ \citenamefont
  {Barab{\'a}si}}]{liu2011controllability}%
  \BibitemOpen
  \bibfield  {author} {\bibinfo {author} {\bibfnamefont {Y.-Y.}\ \bibnamefont
  {Liu}}, \bibinfo {author} {\bibfnamefont {J.-J.}\ \bibnamefont {Slotine}},\
  and\ \bibinfo {author} {\bibfnamefont {A.-L.}\ \bibnamefont {Barab{\'a}si}},\
  }\bibfield  {title} {\bibinfo {title} {Controllability of complex networks},\
  }\href@noop {} {\bibfield  {journal} {\bibinfo  {journal} {Nature}\ }\textbf
  {\bibinfo {volume} {473}},\ \bibinfo {pages} {167} (\bibinfo {year}
  {2011})}\BibitemShut {NoStop}%
\bibitem [{\citenamefont {Menichetti}\ \emph {et~al.}(2014)\citenamefont
  {Menichetti}, \citenamefont {Dall'Asta},\ and\ \citenamefont
  {Bianconi}}]{menichetti2014network}%
  \BibitemOpen
  \bibfield  {author} {\bibinfo {author} {\bibfnamefont {G.}~\bibnamefont
  {Menichetti}}, \bibinfo {author} {\bibfnamefont {L.}~\bibnamefont
  {Dall'Asta}},\ and\ \bibinfo {author} {\bibfnamefont {G.}~\bibnamefont
  {Bianconi}},\ }\bibfield  {title} {\bibinfo {title} {Network controllability
  is determined by the density of low in-degree and out-degree nodes},\
  }\href@noop {} {\bibfield  {journal} {\bibinfo  {journal} {Phys. Rev. Lett.}\
  }\textbf {\bibinfo {volume} {113}},\ \bibinfo {pages} {078701} (\bibinfo
  {year} {2014})}\BibitemShut {NoStop}%
\bibitem [{\citenamefont {Bianconi}\ \emph {et~al.}(2021)\citenamefont
  {Bianconi}, \citenamefont {Sun}, \citenamefont {Rapisardi},\ and\
  \citenamefont {Arenas}}]{bianconi2021message}%
  \BibitemOpen
  \bibfield  {author} {\bibinfo {author} {\bibfnamefont {G.}~\bibnamefont
  {Bianconi}}, \bibinfo {author} {\bibfnamefont {H.}~\bibnamefont {Sun}},
  \bibinfo {author} {\bibfnamefont {G.}~\bibnamefont {Rapisardi}},\ and\
  \bibinfo {author} {\bibfnamefont {A.}~\bibnamefont {Arenas}},\ }\bibfield
  {title} {\bibinfo {title} {Message-passing approach to epidemic tracing and
  mitigation with apps},\ }\href@noop {} {\bibfield  {journal} {\bibinfo
  {journal} {Phys. Rev. Research}\ }\textbf {\bibinfo {volume} {3}},\ \bibinfo
  {pages} {L012014} (\bibinfo {year} {2021})}\BibitemShut {NoStop}%
\bibitem [{\citenamefont {Weigt}\ and\ \citenamefont
  {Zhou}(2006)}]{weigt2006message}%
  \BibitemOpen
  \bibfield  {author} {\bibinfo {author} {\bibfnamefont {M.}~\bibnamefont
  {Weigt}}\ and\ \bibinfo {author} {\bibfnamefont {H.}~\bibnamefont {Zhou}},\
  }\bibfield  {title} {\bibinfo {title} {Message passing for vertex covers},\
  }\href@noop {} {\bibfield  {journal} {\bibinfo  {journal} {Phys. Rev. E}\
  }\textbf {\bibinfo {volume} {74}},\ \bibinfo {pages} {046110} (\bibinfo
  {year} {2006})}\BibitemShut {NoStop}%
\bibitem [{\citenamefont {Dorogovtsev}\ \emph {et~al.}(2008)\citenamefont
  {Dorogovtsev}, \citenamefont {Goltsev},\ and\ \citenamefont
  {Mendes}}]{dorogovtsev2008critical}%
  \BibitemOpen
  \bibfield  {author} {\bibinfo {author} {\bibfnamefont {S.~N.}\ \bibnamefont
  {Dorogovtsev}}, \bibinfo {author} {\bibfnamefont {A.~V.}\ \bibnamefont
  {Goltsev}},\ and\ \bibinfo {author} {\bibfnamefont {J.~F.~F.}\ \bibnamefont
  {Mendes}},\ }\bibfield  {title} {\bibinfo {title} {Critical phenomena in
  complex networks},\ }\href@noop {} {\bibfield  {journal} {\bibinfo  {journal}
  {Rev. Mod. Phys.}\ }\textbf {\bibinfo {volume} {80}},\ \bibinfo {pages}
  {1275} (\bibinfo {year} {2008})}\BibitemShut {NoStop}%
\bibitem [{\citenamefont {Molloy}\ and\ \citenamefont
  {Reed}(1995)}]{molloy1995critical}%
  \BibitemOpen
  \bibfield  {author} {\bibinfo {author} {\bibfnamefont {M.}~\bibnamefont
  {Molloy}}\ and\ \bibinfo {author} {\bibfnamefont {B.}~\bibnamefont {Reed}},\
  }\bibfield  {title} {\bibinfo {title} {A critical point for random graphs
  with a given degree sequence},\ }\href@noop {} {\bibfield  {journal}
  {\bibinfo  {journal} {Random Struct. Algor.}\ }\textbf {\bibinfo {volume}
  {6}},\ \bibinfo {pages} {161} (\bibinfo {year} {1995})}\BibitemShut {NoStop}%
\bibitem [{\citenamefont {Newman}\ \emph {et~al.}(2001)\citenamefont {Newman},
  \citenamefont {Strogatz},\ and\ \citenamefont {Watts}}]{newman2001random}%
  \BibitemOpen
  \bibfield  {author} {\bibinfo {author} {\bibfnamefont {M.~E.~J.}\
  \bibnamefont {Newman}}, \bibinfo {author} {\bibfnamefont {S.~H.}\
  \bibnamefont {Strogatz}},\ and\ \bibinfo {author} {\bibfnamefont {D.~J.}\
  \bibnamefont {Watts}},\ }\bibfield  {title} {\bibinfo {title} {Random graphs
  with arbitrary degree distributions and their applications},\ }\href@noop {}
  {\bibfield  {journal} {\bibinfo  {journal} {Phys. Rev. E}\ }\textbf {\bibinfo
  {volume} {64}},\ \bibinfo {pages} {026118} (\bibinfo {year}
  {2001})}\BibitemShut {NoStop}%
\bibitem [{\citenamefont {Newman}(2009)}]{newman2009random}%
  \BibitemOpen
  \bibfield  {author} {\bibinfo {author} {\bibfnamefont {M.~E.~J.}\
  \bibnamefont {Newman}},\ }\bibfield  {title} {\bibinfo {title} {Random graphs
  with clustering},\ }\href@noop {} {\bibfield  {journal} {\bibinfo  {journal}
  {Phys. Rev. Lett.}\ }\textbf {\bibinfo {volume} {103}},\ \bibinfo {pages}
  {058701} (\bibinfo {year} {2009})}\BibitemShut {NoStop}%
\bibitem [{\citenamefont {Cho}\ \emph {et~al.}(2009)\citenamefont {Cho},
  \citenamefont {Kim}, \citenamefont {Park}, \citenamefont {Kahng},\ and\
  \citenamefont {Kim}}]{kahng009percolation}%
  \BibitemOpen
  \bibfield  {author} {\bibinfo {author} {\bibfnamefont {Y.~S.}\ \bibnamefont
  {Cho}}, \bibinfo {author} {\bibfnamefont {J.~S.}\ \bibnamefont {Kim}},
  \bibinfo {author} {\bibfnamefont {J.}~\bibnamefont {Park}}, \bibinfo {author}
  {\bibfnamefont {B.}~\bibnamefont {Kahng}},\ and\ \bibinfo {author}
  {\bibfnamefont {D.}~\bibnamefont {Kim}},\ }\bibfield  {title} {\bibinfo
  {title} {{Percolation transitions in scale-free networks under the Achlioptas
  process}},\ }\href@noop {} {\bibfield  {journal} {\bibinfo  {journal} {Phys.
  Rev. Lett.}\ }\textbf {\bibinfo {volume} {103}},\ \bibinfo {pages} {135702}
  (\bibinfo {year} {2009})}\BibitemShut {NoStop}%
\bibitem [{\citenamefont {Li}\ \emph {et~al.}(2021)\citenamefont {Li},
  \citenamefont {Liu}, \citenamefont {L{\"u}}, \citenamefont {Hu},
  \citenamefont {Xu},\ and\ \citenamefont {Zhang}}]{li2021percolation}%
  \BibitemOpen
  \bibfield  {author} {\bibinfo {author} {\bibfnamefont {M.}~\bibnamefont
  {Li}}, \bibinfo {author} {\bibfnamefont {R.-R.}\ \bibnamefont {Liu}},
  \bibinfo {author} {\bibfnamefont {L.}~\bibnamefont {L{\"u}}}, \bibinfo
  {author} {\bibfnamefont {M.-B.}\ \bibnamefont {Hu}}, \bibinfo {author}
  {\bibfnamefont {S.}~\bibnamefont {Xu}},\ and\ \bibinfo {author}
  {\bibfnamefont {Y.-C.}\ \bibnamefont {Zhang}},\ }\bibfield  {title} {\bibinfo
  {title} {{Percolation on complex networks: Theory and application}},\
  }\href@noop {} {\bibfield  {journal} {\bibinfo  {journal} {Phys. Rep.}\
  }\textbf {\bibinfo {volume} {907}},\ \bibinfo {pages} {1} (\bibinfo {year}
  {2021})}\BibitemShut {NoStop}%
\bibitem [{\citenamefont {Lee}\ \emph {et~al.}(2018)\citenamefont {Lee},
  \citenamefont {Kahng}, \citenamefont {Cho}, \citenamefont {Goh},\ and\
  \citenamefont {Lee}}]{lee2018recent}%
  \BibitemOpen
  \bibfield  {author} {\bibinfo {author} {\bibfnamefont {D.}~\bibnamefont
  {Lee}}, \bibinfo {author} {\bibfnamefont {B.}~\bibnamefont {Kahng}}, \bibinfo
  {author} {\bibfnamefont {Y.~S.}\ \bibnamefont {Cho}}, \bibinfo {author}
  {\bibfnamefont {K.-I.}\ \bibnamefont {Goh}},\ and\ \bibinfo {author}
  {\bibfnamefont {D.-S.}\ \bibnamefont {Lee}},\ }\bibfield  {title} {\bibinfo
  {title} {Recent advances of percolation theory in complex networks},\
  }\href@noop {} {\bibfield  {journal} {\bibinfo  {journal} {J. Korean Phys.
  Soc.}\ }\textbf {\bibinfo {volume} {73}},\ \bibinfo {pages} {152} (\bibinfo
  {year} {2018})}\BibitemShut {NoStop}%
\bibitem [{\citenamefont {Dorogovtsev}\ and\ \citenamefont
  {Mendes}(2022)}]{dorogovtsev2022nature}%
  \BibitemOpen
  \bibfield  {author} {\bibinfo {author} {\bibfnamefont {S.~N.}\ \bibnamefont
  {Dorogovtsev}}\ and\ \bibinfo {author} {\bibfnamefont {J.~F.~F.}\
  \bibnamefont {Mendes}},\ }\href@noop {} {\emph {\bibinfo {title} {The Nature
  of Complex Networks}}}\ (\bibinfo  {publisher} {Oxford University Press,
  Oxford},\ \bibinfo {year} {2022})\BibitemShut {NoStop}%
\bibitem [{\citenamefont {Ferraz~de Arruda}\ \emph {et~al.}(2021)\citenamefont
  {Ferraz~de Arruda}, \citenamefont {Tizzani},\ and\ \citenamefont
  {Moreno}}]{ferraz2021phase}%
  \BibitemOpen
  \bibfield  {author} {\bibinfo {author} {\bibfnamefont {G.}~\bibnamefont
  {Ferraz~de Arruda}}, \bibinfo {author} {\bibfnamefont {M.}~\bibnamefont
  {Tizzani}},\ and\ \bibinfo {author} {\bibfnamefont {Y.}~\bibnamefont
  {Moreno}},\ }\bibfield  {title} {\bibinfo {title} {Phase transitions and
  stability of dynamical processes on hypergraphs},\ }\href@noop {} {\bibfield
  {journal} {\bibinfo  {journal} {Commun. Phys.}\ }\textbf {\bibinfo {volume}
  {4}},\ \bibinfo {pages} {1} (\bibinfo {year} {2021})}\BibitemShut {NoStop}%
\bibitem [{\citenamefont {Chodrow}\ \emph {et~al.}(2021)\citenamefont
  {Chodrow}, \citenamefont {Veldt},\ and\ \citenamefont
  {Benson}}]{chodrow2021generative}%
  \BibitemOpen
  \bibfield  {author} {\bibinfo {author} {\bibfnamefont {P.~S.}\ \bibnamefont
  {Chodrow}}, \bibinfo {author} {\bibfnamefont {N.}~\bibnamefont {Veldt}},\
  and\ \bibinfo {author} {\bibfnamefont {A.~R.}\ \bibnamefont {Benson}},\
  }\bibfield  {title} {\bibinfo {title} {{Generative hypergraph clustering:
  From blockmodels to modularity}},\ }\href@noop {} {\bibfield  {journal}
  {\bibinfo  {journal} {Sci. Adv.}\ }\textbf {\bibinfo {volume} {7}},\ \bibinfo
  {pages} {eabh1303} (\bibinfo {year} {2021})}\BibitemShut {NoStop}%
\bibitem [{\citenamefont {Stewart~III}\ and\ \citenamefont
  {Woon}(2008)}]{stewart2008congressional}%
  \BibitemOpen
  \bibfield  {author} {\bibinfo {author} {\bibfnamefont {C.}~\bibnamefont
  {Stewart~III}}\ and\ \bibinfo {author} {\bibfnamefont {J.}~\bibnamefont
  {Woon}},\ }\href@noop {} {\emph {\bibinfo {title} {Congressional committee
  assignments, 103rd to 114th congresses, 1993--2017: House}}},\ \bibinfo
  {type} {Tech. Rep.}\ (\bibinfo  {institution} {MIT mimeo},\ \bibinfo {year}
  {2008})\BibitemShut {NoStop}%
\end{thebibliography}%

%%\newpage

%%%%%%%%%%%%%%%%%%%%%%%%%%%%%%%%%%%%%%%%%%%%%%%%%%%
%%%%%%%%%%%%%%%%%%%%%%%%%%%%%%%%%%%%%%%%%%%%%%%%%%%
%%%%%%%%%%%%%%%%%%%%%%%%%%%%%%%%%%%%%%%%%%%%%%%%%%%
%%%%%%%%%%%%%%%%%%%%%%%%%%%%%%%%%%%%%%%%%%%%%%%%%%%

\end{document}